%% file: arxiv.tex
\documentclass[11pt]{article}

\usepackage[letterpaper,margin=1in]{geometry}
\usepackage[T1]{fontenc}
\usepackage{microtype}
\usepackage{newtxtext,newtxmath} 

\usepackage[hidelinks]{hyperref}
\hypersetup{
  pdftitle={Who Owns the Text? Design Patterns for Preserving Authorship in AI-Assisted Writing},
  pdfauthor={Bohan Zhang, Chengke Bu, Paramveer Dhillon},
}

\usepackage[numbers,sort&compress]{natbib}

\usepackage{graphicx}
\usepackage{subcaption}
\usepackage{booktabs}
\usepackage{longtable}
\usepackage{float}
\usepackage{xcolor}
\usepackage{tikz}
\usepackage[most]{tcolorbox}
\usepackage{pifont}
\usepackage[normalem]{ulem}
\usepackage{enumitem}
\setlist[itemize]{nosep,leftmargin=*}
\setlist[enumerate]{nosep,leftmargin=*}
\providecommand{\Description}[1]{}
\newcommand{\add}[1]{#1}
\newcommand{\del}[1]{}

\newcommand{\keywords}[1]{\par\smallskip\noindent\textbf{Keywords:} #1\par}

\usepackage{authblk}
\author[1]{Bohan Zhang}
\author[2]{Chengke Bu}
\author[1]{Paramveer Dhillon\thanks{Corresponding author: \texttt{dhillonp@umich.edu}}}

\affil[1]{University of Michigan, Ann Arbor, Michigan, USA}
\affil[2]{Tsinghua University, Beijing, China}

\date{} 

\title{Who Owns the Text? Design Patterns for Preserving Authorship in AI-Assisted Writing}

\begin{document}
\maketitle

\begin{abstract}
AI writing assistants can reduce effort and improve fluency, but they may also weaken writers’ sense of authorship. We study this tension with an ownership-aware co-writing editor that offers on-demand, sentence-level suggestions and tests two common design choices: persona-based coaching and style personalization. In an online study ($N{=}176$), participants completed three professional writing tasks—an email without AI help, a proposal with generic AI suggestions, and a cover letter with persona-based coaching—while half received suggestions tailored to a brief sample of their prior writing. Across the two AI-assisted tasks, psychological ownership dropped relative to unassisted writing (about 0.85–1.0 points on a 7-point scale), even as cognitive load decreased (about 0.9 points) and quality ratings stayed broadly similar overall. Persona coaching did not prevent the ownership decline. Style personalization partially restored ownership (about +0.43) and increased AI incorporation in text (+5 percentage points). We distill five design patterns—on-demand initiation, micro-suggestions, voice anchoring, audience scaffolds, and point-of-decision provenance—to guide authorship-preserving writing tools.
\end{abstract}



\maketitle

\section{Introduction}
Large language models (LLMs) have reshaped writing by making high-quality drafting, continuation, and revision available on demand. Systems such as GPT now power suggestion and rewriting features in consumer and workplace tools, and HCI research has explored their use in creative co-authoring, collaborative story generation, screenwriting support, and newsroom workflows \citep{brown2020language,buschek2021impact,clark2018creative,yuan2022wordcraft,mirowski2023co,diakopoulos2019automating,dorr2016mapping,kim2024diarymate,qin2024charactermeet}. As these systems move from standalone chat into writing interfaces, designers increasingly make choices about \emph{how} assistance is framed and delivered, for example through persona-framed coaching or style-personalized suggestions \citep{lee2024design,li2024value}. These design choices can plausibly shift not only productivity and quality, but also how writers experience authorship \citep{zhao2025making}.

Integrating LLMs into writing raises questions about agency and psychological ownership: whether writers feel the text is “mine” when suggestions are fluent, persuasive, and easy to adopt. Prior studies show that co-writing with AI can reduce psychological ownership, and that increasing writer investment (for example, requiring longer prompts) can partially mitigate this loss \citep{joshi2025writing}. Related work also suggests that perceived inclusion and control shape agency and ownership in AI-mediated communication \citep{kadoma2024role}, and that model behavior can steer users’ judgments and viewpoints \citep{jakesch2023co}. These concerns parallel broader observations that AI reconfigures work practices and role boundaries \citep{seeber2020machines}. Together, this literature motivates examining two design levers that are now common in deployed writing tools: persona framing, which can scaffold audience and tone through a role-based assistant, and style personalization, which can align suggestions with a writer’s habitual phrasing \citep{lee2024design,li2024value,zhao2025making}. A reasonable expectation from prior work is that persona framing may increase engagement and audience awareness without necessarily transferring authorship, whereas style personalization may better support “voice fit” and thereby strengthen psychological ownership \citep{benharrak2024writer,qin2025toward,draxler2024ai,gero2023social}.

This creates a design tension: AI assistance can reduce effort and improve fluency while also weakening felt authorship and authorial identity. Outside academia, authorship debates are already visible in public discourse around AI-assisted writing, including high-profile cases such as Rie Kudan’s comments about using ChatGPT \citep{choi2024winner}. More broadly, writers’ ownership concerns extend beyond the immediate drafting experience to questions about how writing is used as training data and how authorship is understood in AI-mediated ecosystems \citep{gero2025creative}. For professional writing in particular, the design challenge is not whether LLMs can produce competent text, but how interfaces can preserve user agency and autonomy while still leveraging AI support \citep{hancock2020ai,luger2016like,bhat2024llms,lim2024co,zhou2024glassmail}.

AI writing tools are also not merely interfaces we interact with; they are systems we entangle our identities with. When a writer accepts an AI suggestion, the question “who wrote this?” becomes not just a matter of attribution but of self-presentation and professional identity \citep{zhao2025making,gero2025creative}. This positions ownership not as a psychological curiosity, but as a design problem that extends beyond discrete interaction events into ongoing questions of voice, accountability, and creative selfhood.

In this paper, we ask:
\begin{enumerate}
\item \textbf{How do persona framing and style personalization\textemdash individually and in combination\textemdash shape writers’ \emph{psychological ownership} of AI-assisted text?} In particular, does style personalization mitigate ownership loss, and does it matter more in audience-facing writing where persona framing is commonly used?
\item \textbf{What are the downstream consequences for the writing process\textemdash perceived workload/effort and perceived quality\textemdash under genre-appropriate assistance modes?}
\end{enumerate}

We study these questions through two interface-level levers. First, we examine \emph{persona-framed coaching}, where the assistant is presented as a role-based helper with an explicit purpose and voice. In our system, “Emma” is positioned as a style-and-tone coach who offers short, adoptable suggestions for audience-facing writing. Prior work suggests that social and framing choices can shape how writers interpret and relate to AI support in writing contexts \citep{chaves2021should,seeger2021texting,gero2023social}. Second, we examine \emph{style personalization}, where the assistant conditions its suggestions on a brief sample of the writer’s prior text. This lever is motivated by self-extension accounts of ownership \citep{belk1988possessions} and by writing-assistance research showing that tailoring support can improve fit and user experience \citep{lee2022coauthor,benharrak2024writer,qin2025toward,zhao2025making}. Rather than treating personas and personalization as alternatives, we treat them as complementary: personas can scaffold audience and tone, while personalization can anchor voice, potentially making adopted text feel more like it belongs to the writer \citep{lee2024design,li2024value}.

\begin{figure}[htbp]
    \centering
    \includegraphics[width=0.9\linewidth]{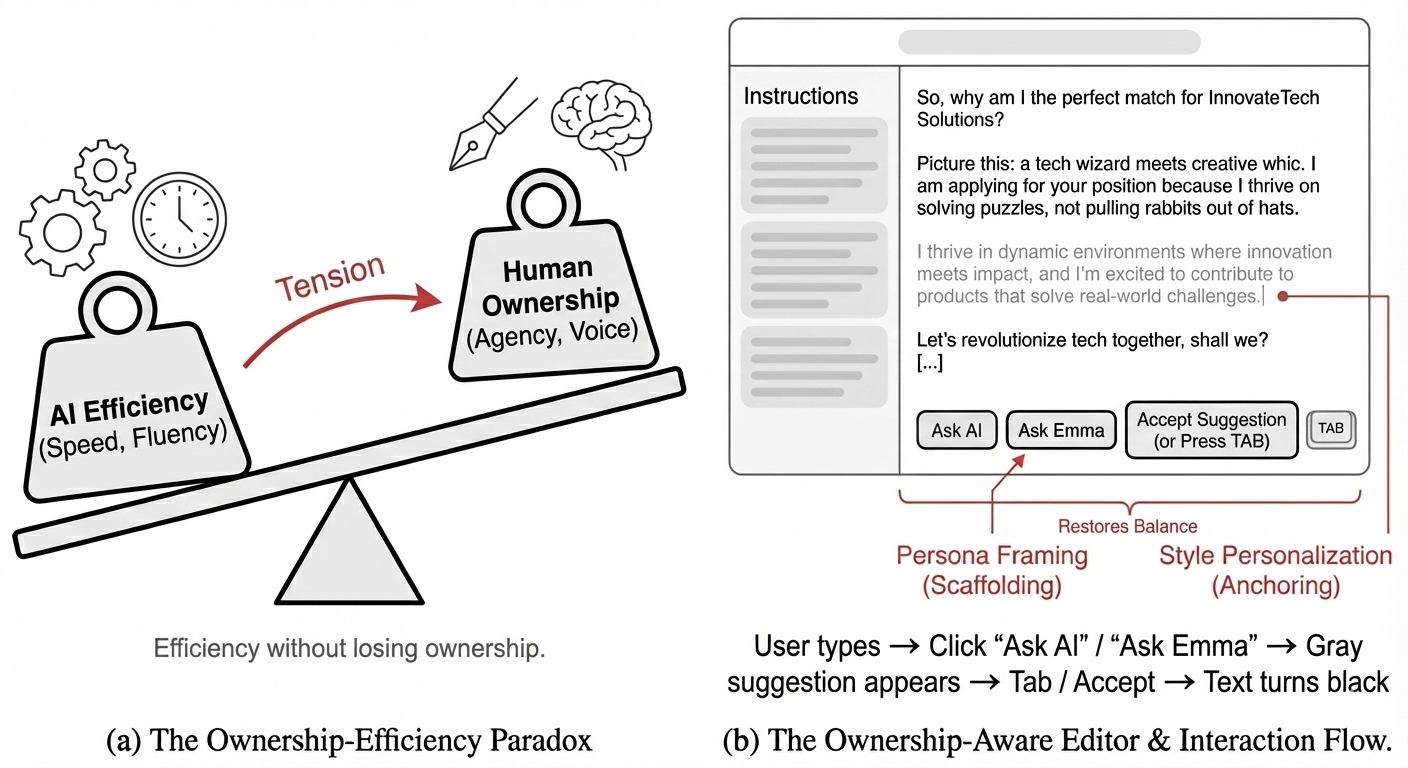}
    \caption{\textbf{The ownership--efficiency paradox and our ownership-aware editor.}
    (a) AI writing assistance can increase efficiency (speed, fluency/quality) while reducing human ownership (agency, voice), creating an ownership--efficiency tension.
    (b) Our editor operationalizes two levers within an on-demand micro-suggestion loop: writers click \textit{Ask Emma} (persona-framed coaching) or \textit{Ask AI} (generic suggestions), a gray inline suggestion appears, and writers explicitly accept (Tab/button) or ignore it; accepted text turns black and remains editable.}
     \Description{Two-panel figure. (a) A seesaw schematic contrasts AI efficiency (speed, fluency/quality) with human ownership (agency, voice), labeled as an ownership–efficiency tension. (b) A co-writing editor mock shows “Ask AI” and “Ask Emma” buttons; a gray inline suggestion appears in the draft and can be accepted via Tab/button to turn black and remain editable, or ignored.}
    \label{fig:paradox_editor}
\end{figure}

To evaluate these levers in a realistic professional-writing context, we conducted an IRB-approved study with 176 Prolific participants. Each participant completed three genre-specific writing scenarios in a within-subject design: an \emph{unassisted email baseline} (no AI), a \emph{generic AI suggestion mode for proposals} (contextual next-sentence continuations without persona framing), and a \emph{persona-framed coaching mode for cover letters} (Emma). \emph{Style personalization} was manipulated between subjects by conditioning suggestions on a prior writing sample. We adopt a matched-task setup to reflect the practical reality that assistance modes are often genre-sensitive rather than one-size-fits-all \citep{lee2024design,zhao2025making}: cover letters are explicitly audience-facing and lend themselves to coaching-style framing, proposals benefit from contextual continuation, and the unassisted email provides a baseline for ownership and effort. Scenario order was counterbalanced using a Latin square design, and we focus our analyses on within-genre comparisons rather than pooling across dissimilar writing tasks. Figure~\ref{fig:paradox_editor} situates the work in the ownership--efficiency tension and previews the editor interaction flow.

Our findings show that both assisted scenarios—generic suggestions and persona-framed coaching—reduced writers’ psychological ownership relative to the unassisted baseline (about 1.0 and 0.85 points on a 7-point scale, respectively). This ownership reduction persisted even when assistance was delivered through a persona-framed coaching interface. Style personalization, however, partially restored ownership (about +0.43 points), suggesting that improving “voice fit” plays a more direct role in preserving authorship than framing alone. Across scenarios, self-reported workload decreased under AI assistance, while evaluator-rated quality was broadly similar overall, with a modest decrease in the generic suggestion/ configuration. Personalization also increased adoption, with participants incorporating about five percentage points more suggested text. Together, these results highlight an ownership–efficiency tension: AI assistance can reduce effort without harming output in a large way, yet still weaken the sense that the text is one’s own.

This paper makes three contributions. First, we present an interaction design for professional AI co-writing that pairs persona framing (audience/tone scaffold) with lightweight style personalization (voice anchoring) using bounded micro-suggestions and provenance cues \citep{lee2024design,hoque2024hallmark,buschek2024collage}. Second, we provide evaluation evidence ($N{=}176$) on how these design choices relate to psychological ownership, workload, and perceived quality in genre-appropriate writing scenarios. Third, we distill five design patterns, with implementation guidance and boundary conditions, to inform the design of AI writing tools that support productivity without treating authorship as collateral damage.

\section{Background and Design Rationale}

\subsection{Psychological ownership as a design concern in digital creation}
Psychological ownership describes the feeling that something is ``mine,'' independent of legal ownership \citep{pierce2001toward}. Although originally developed in organizational behavior, the construct has proven useful for understanding how people relate to digital artifacts that are authored, curated, and publicly displayed. Prior work finds that users can experience ``mine-ness'' toward social profiles, posts, and other user-generated content \citep{karahanna2015psychological,sinclair2017psychological}, as well as toward virtual goods in games and online environments \citep{guo2012explaining}. In these settings, ownership is not only about possession of an object. It is also tied to identity and self-presentation: digital artifacts often stand in for the self, and people can feel ownership over the voice, style, and audience relationship that a platform enables \citep{gero2025creative}.

Collaboration complicates these dynamics because authorship becomes distributed across contributors and tools. Studies of Wikipedia and document co-authoring show that people negotiate boundaries of contribution, territory, and shared ownership norms in jointly produced texts \citep{thom2009s,blau2009type}. These negotiations are not merely procedural; they shape whether contributors feel pride, responsibility, and attachment to what is produced. AI-assisted writing adds a new kind of collaborator that can generate content quickly and persuasively. As a result, psychological ownership becomes a practical design concern: if users systematically feel less attachment and responsibility for text produced with AI assistance, then writing tools may trade away authorship in exchange for efficiency \citep{li2024value,zhao2025making}.

\subsection{AI-assisted writing as an interface design space}
LLM-based writing support is now deployed across consumer and professional contexts, including creative co-authoring, story generation, personal journaling, and professional workflows \citep{buschek2021impact,yuan2022wordcraft,mirowski2023co,kim2024diarymate,qin2024charactermeet,bhat2024llms,lim2024co,zhou2024glassmail}. From a design perspective, these systems do more than generate fluent text. They offer suggestions, provide alternatives, restructure passages, and help users explore phrasing, tone, and rhetorical stance \citep{lee2024design,li2024value,buschek2024collage}. The resulting user experience depends strongly on interaction details: when suggestions appear, how much text is proposed at once, how suggestions are accepted or revised, and how the system communicates its role in the writing process \citep{lee2024design,zhao2025making}.

Prior work suggests that these systems can change both process and experience. LLM assistance has been associated with lower perceived effort and shifts in how writers plan and revise \citep{lee2022coauthor,li2024value}. At the same time, the presence of an AI generator can complicate authenticity and authorship: users may perceive output as less ``theirs,'' even when they initiated the request and made final decisions \citep{joshi2025writing,draxler2024ai,gero2025creative}. In related work on AI-mediated communication, model behavior can also steer users' judgments and viewpoints \citep{jakesch2023co}, underscoring that writing support is not neutral even when it is framed as ``help.'' These observations motivate a design goal that goes beyond producing high-quality text: writing interfaces should support efficiency while preserving agency and authorship \citep{lee2024design,li2024value,zhao2025making}.

\subsection{Agency, control, and the legibility of contribution in human--AI co-writing}
A recurring theme in human--AI interaction is that user agency depends on both capability and interface design. Guidelines for human-centered AI emphasize supporting meaningful user control, enabling oversight, and designing interactions where users can accept, reject, or revise system output in ways that preserve autonomy \citep{amershi2019guidelines}. Work on control and adjustable autonomy similarly argues that interfaces should allow people to tune how much initiative and influence an automated system has in a task \citep{feigh2014requirements,wang2019designing}. In writing contexts, these principles translate into concrete design questions: Should the system proactively suggest text or wait for explicit requests? Should it propose long passages or short increments? How visible should AI contributions be during writing? How easily can users undo or reshape adopted suggestions \citep{lee2024design,dhillon2024shaping}?

Agency is also shaped by cognitive factors. The timing, granularity, and presentation of suggestions can steer decisions and perceived control \citep{lai2022human,green2019principles}. For example, even short suggestions can change planning and voice, especially when they are fluent and low-friction to accept \citep{bhat2023interacting,buschek2024collage}. From an ownership perspective, this matters because control is one pathway through which mine-ness develops: people are more likely to feel ownership over outcomes they intentionally shaped. Consistent with this, prior work finds that increasing writer investment, such as requiring more explicit input during prompting, can partially mitigate ownership loss in AI-assisted drafting \citep{joshi2025writing}. This suggests that ownership is not fixed by the presence of AI alone; it is sensitive to interaction design choices that determine how much the user directs, edits, and integrates suggestions \citep{lee2024design,zhao2025making}.

A second, closely related issue is the legibility of contribution. When AI output blends seamlessly into a draft, users may have difficulty tracking what originated from them versus what originated from the system. Work on attribution and credit in human--AI co-creation suggests that perceived credit is shaped by contribution and control rather than by superficial framing \citep{he2025contributions}. This aligns with practical concerns in writing: when the system's influence becomes hard to perceive, users may experience reduced ownership, reduced responsibility, or uncertainty about what to disclose. Even when a tool improves fluency, it may still produce a ``ghostwriter'' effect in which the text feels authored by the system rather than by the user \citep{draxler2024ai}. Recent writing-focused systems also emphasize provenance and transparency as design requirements for responsible use of LLMs in writing \citep{hoque2024hallmark}. Together, these lines of work point to a design imperative: AI writing interfaces should make contribution and influence legible enough that users can maintain agency and authorship judgments \citep{hoque2024hallmark,buschek2024collage}.

\subsection{Two design levers for preserving authorship: persona framing and style personalization}
Within this design space, we focus on two levers that are now common in deployed writing tools and that map cleanly onto authorship concerns: persona framing and style personalization \citep{lee2024design,zhao2025making}.

\textbf{Persona framing.} Persona-framed assistants present the system as a consistent role with a purpose, stance, and communication style. In writing tools, personas often function as audience and tone scaffolds: they can help writers hold an intended reader in mind and interpret suggestions as coaching rather than as ghostwriting. Prior HCI work on anthropomorphic and conversational agents suggests that framing can shape engagement and how users interpret system output \citep{chaves2021should,seeger2021texting}. In creative writing, social dynamics and framing can shape how writers interpret and negotiate AI support \citep{gero2023social}. In creative and writing settings, persona-like scaffolds have been used to support exploration and writer goals \citep{benharrak2024writer,qin2024charactermeet}. However, existing evidence also suggests that framing alone may not transfer authorship: if control and contribution remain dominated by the system, users may still experience reduced ownership even when the assistant appears helpful or socially present \citep{he2025contributions,draxler2024ai}.

\textbf{Style personalization.} Style personalization adapts suggestions to a writer's habitual phrasing and structure, aiming to increase the perceived fit of AI output. This lever is motivated by self-extension accounts of ownership, which argue that possessions and creations can become part of the self \citep{belk1988possessions}. If adopted suggestions align with a writer's voice, they may feel more self-consistent and therefore more ``mine.'' In writing support, personalization has been proposed as a way to keep agency with the writer while benefiting from generation \citep{lee2022coauthor,qin2025toward,zhao2025making}. At the same time, personalization raises practical tradeoffs, including privacy concerns (it requires collecting writing samples) and the possibility that the system's style model is imperfect or overly constraining. More broadly, recent work suggests that writing with language models can reduce content diversity and contribute to homogenization, raising additional concerns about convergence toward generic defaults \citep{padmakumar2023does,li2024value}. From a design standpoint, the key question is whether lightweight personalization can improve authorship experience without increasing burden or reducing the utility of suggestions.

\subsection{Design stance and rationale}
This background motivates a design stance that is central to our study: persona framing and style personalization may play complementary roles in AI-assisted professional writing. Personas can scaffold audience and tone by framing the assistant as a coach rather than a co-author, while personalization can anchor voice by aligning suggestions with the writer's habitual style. Importantly, neither lever operates in isolation from interface choices that shape agency and legibility, such as on-demand initiation, bounded suggestion length, and provenance cues \citep{lee2024design,hoque2024hallmark,buschek2024collage}. Our work therefore evaluates persona framing and style personalization as practical levers in a professional writing context, focusing on how they shape psychological ownership alongside workload and perceived quality, and using the findings to motivate design patterns for authorship-preserving AI writing tools \citep{zhao2025making}.

\section{System: An Ownership-Aware AI Co-Writing Editor}
\label{sec:system}

We built a browser-based co-writing editor to study how interface-level design choices in AI writing assistance shape writers’ sense of authorship. The system instantiates two levers that are increasingly common in deployed tools: \emph{persona framing} (a role-based coaching assistant) and \emph{style personalization} (conditioning suggestions on a writer’s prior text) \citep{lee2024design,li2024value}. The editor is designed around a lightweight interaction model—on-demand, sentence-level suggestions that are visually distinguishable and easy to accept or ignore—so that writers retain control over when and how AI enters the writing process \citep{lee2024design,buschek2024collage}.

\subsection{Design rationale}
\label{sec:system-rationale}
Our design choices were guided by three practical constraints. First, we needed an interaction model that enables a controlled evaluation of psychological ownership without requiring participants to learn a complex workflow. Second, the system had to preserve ecological validity in professional writing, where writers often request help at moments of friction (e.g., phrasing a sentence, adjusting tone) rather than delegating the entire draft \citep{lee2024design,li2024value}. Third, we aimed to reduce the risk that the model would overwrite a participant’s voice through long completions or opaque blending of AI text into the draft. These constraints motivated four interaction decisions that align with recent accounts of writing assistants as an interface design space \citep{lee2024design,li2024value}:

\begin{itemize}
    \item \textbf{On-demand initiation.} The system generates suggestions only when writers explicitly request them, preserving agency over when assistance enters the workflow \citep{lee2024design}.
    \item \textbf{Bounded micro-suggestions.} Suggestions are short and adoptable (one sentence; capped at 30 tokens), limiting the chance of extended AI takeovers and supporting incremental revision \citep{buschek2024collage}.
    \item \textbf{Visual provenance at the point of decision.} Suggestions appear in light gray and are visually distinct from user-authored text while being considered; accepted text turns black, matching the user’s draft \citep{hoque2024hallmark}.
    \item \textbf{Lightweight style personalization.} For writers in the personalized condition, suggestions are conditioned on a brief writing sample to increase stylistic fit without requiring extensive configuration \citep{lee2024design,li2024value}.
\end{itemize}

Together, these decisions position the system as a co-writing tool that supports efficiency while making the system’s influence legible during suggestion adoption and leaving final authorship decisions with the writer.

\subsection{Interface and interaction model}
\label{sec:system-interaction}
The editor presents task instructions on the left and a writing window on the right. Figure~\ref{fig:paradox_editor}(b) illustrates the editor interface and the request$\rightarrow$preview$\rightarrow$accept/ignore interaction flow; additional screenshots appear in Appendix~\ref{app:interface}. We implemented the writing window using the Quill rich-text editor to support standard operations (typing, selection, cursor navigation, paragraphing). A live word count and timer are displayed, and submission is enabled only after the draft reaches a 200-word threshold, ensuring comparable engagement across tasks.

In the assisted scenarios, writers can request an inline suggestion via a single button. The button label reflects the assistance mode: \textit{Ask AI} for generic assistance and \textit{Ask Emma} for persona-framed coaching. After a request, the system inserts a suggestion at the cursor in light gray, visually separating it from the writer’s black text. Writers can accept the suggestion up to the next punctuation mark using the \textit{Tab} key or an \textit{Accept Suggestion} button; accepted text becomes black and remains editable like the rest of the draft. Writers can also ignore the suggestion and continue typing. Each suggestion is bounded to a single sentence and capped to limit takeover and support incremental incorporation. Full API parameters are provided in Appendix~\ref{app:api}. This interaction style reflects a broader trend toward fragmentary, edit-and-assemble workflows in AI writing tools \citep{buschek2024collage}.

This interaction model—request \(\rightarrow\) preview \(\rightarrow\) accept/ignore—was chosen to keep assistance easy to invoke and easy to reject. It also aligns with the goal of preserving agency: writers determine when to request help and which parts of a suggestion to adopt \citep{lee2024design}.

\subsection{Assistance modes: generic AI suggestions and persona-framed coaching}
\label{sec:system-modes}
The system supports two AI-assisted modes that share the same basic mechanism (context-aware next-sentence suggestions) but differ in framing and guidance.

\textbf{Generic AI suggestion mode.} In the generic mode, the model provides a context-aware continuation based on the current draft without adopting a persona or an explicit coaching role. The prompt emphasizes producing a plausible next sentence that follows the writer’s content.

\textbf{Persona-framed coaching mode.} In the persona-framed mode, the assistant is presented as ``Emma,'' a style-and-tone coach for cover-letter writing. Emma’s role is explicitly described as helping the writer refine their draft—tightening phrasing, improving tone, and strengthening audience fit—rather than producing content on the writer’s behalf. This framing is intended to position suggestions as coaching-oriented refinements rather than as ghostwritten text. Full prompt templates for both modes appear in Appendix~\ref{app:prompts}.

Importantly, these modes differ in how assistance is framed and instructed, while holding constant the core interaction pattern (on-demand, sentence-level suggestions) so that we can examine how role framing interacts with writers’ experience of authorship.

\subsection{Style personalization pipeline}
\label{sec:system-personalization}
To study whether stylistic fit affects ownership and adoption, the system optionally personalizes suggestions using a brief sample of the writer’s prior text. Writers assigned to the personalized condition submit a short writing sample before beginning the assisted tasks. The system uses GPT-4o to extract a compact set of style descriptors (five characteristics such as tone, formality, sentence structure, lexical choice, and rhetorical tendencies). These descriptors are injected into the suggestion prompts used in the assisted modes, instructing the model to align its suggestions with the writer’s habitual voice \citep{lee2024design,li2024value}. Writers in the non-personalized group use the same interface and assistance modes, but without any style-conditioning text.

This ``lightweight profile'' approach is intended to keep personalization feasible in practical tools (where users may not want to configure many settings) while still producing measurable differences in perceived stylistic fit.

\subsection{Implementation details and telemetry}
\label{sec:system-implementation}
The system uses a Node.js backend with an HTML/CSS/JavaScript frontend and was hosted on Fly.io. AI assistance was provided by GPT-4o.\footnote{\url{https://openai.com/index/hello-gpt-4o/}} The editor records the full task text and a detailed interaction log, including suggestion requests, accept/ignore events, timestamps, word counts, and keystroke-derived edit activity. These logs support analyses of writing process measures (e.g., time, edits) and adoption behavior (e.g., the fraction of AI-generated text incorporated into the final draft), which we relate to self-reported outcomes in subsequent sections.

\paragraph{Telemetry and provenance measurement.}
While provenance cues in the interface are primarily designed to make suggestions legible at the point of decision (gray preview text) \citep{hoque2024hallmark}, the system also logs acceptance events and the inserted suggestion content. This enables an estimate of the proportion of AI-generated text incorporated into the final draft, which we use as a behavioral complement to self-reported measures of ownership and workload.

\section{Evaluation Approach}
\label{sec:evaluation}

We evaluate the system by asking how two design levers—persona framing and style personalization—shape writers’ psychological ownership and writing experience in professional contexts. Our evaluation is intentionally \emph{genre-sensitive}: rather than applying one assistance mode to every genre, we pair each assistance mode with a professional writing scenario where it plausibly fits. Figure~\ref{fig:style-interface} summarizes the three scenarios and the interface.

\subsection{Study design and tasks}
\label{sec:study-design}
We used a mixed design with one within-subject factor (writing scenario / assistance mode) and one between-subject factor (style personalization). Each participant completed three writing scenarios (within subjects), each requiring at least 200 words:

\begin{itemize}
    \item \textbf{Scenario A: Unassisted email baseline.} Participants drafted a professional email proposing a software-integration partnership with no AI assistance. This scenario establishes baseline ownership and effort without AI tool influence.
    \item \textbf{Scenario B: Generic suggestion mode for proposals.} Participants wrote a project-feature proposal with access to contextual next-sentence suggestions generated by the model, without persona framing.
    \item \textbf{Scenario C: Persona-framed coaching mode for cover letters.} Participants wrote a cover letter with support from ``Emma,'' a role-based style-and-tone coach offering short, adoptable micro-suggestions.
\end{itemize}

This matched-task setup reflects our position that AI writing assistance should be genre-sensitive rather than one-size-fits-all. In professional practice, writers often seek different kinds of support depending on the audience and rhetorical demands of the genre: cover letters are explicitly audience-facing and can benefit from coaching-style tone scaffolds, proposals often benefit from contextual continuation that helps elaborate ideas, and an unassisted email provides a natural reference point for ownership and workload. The scenarios share a common domain (professional technology communication) while differing in genre and rhetorical goals, reducing direct content reuse and limiting strategy carryover across tasks.

Task order was counterbalanced using a Latin square. Participants were randomly assigned to one of three sequences, and we include sequence position as a control in analysis. Personalization was manipulated between subjects: half of participants provided a writing sample used to condition AI suggestions in the two assisted scenarios.

\begin{figure}[t]
  \centering
  \includegraphics[width=\linewidth]{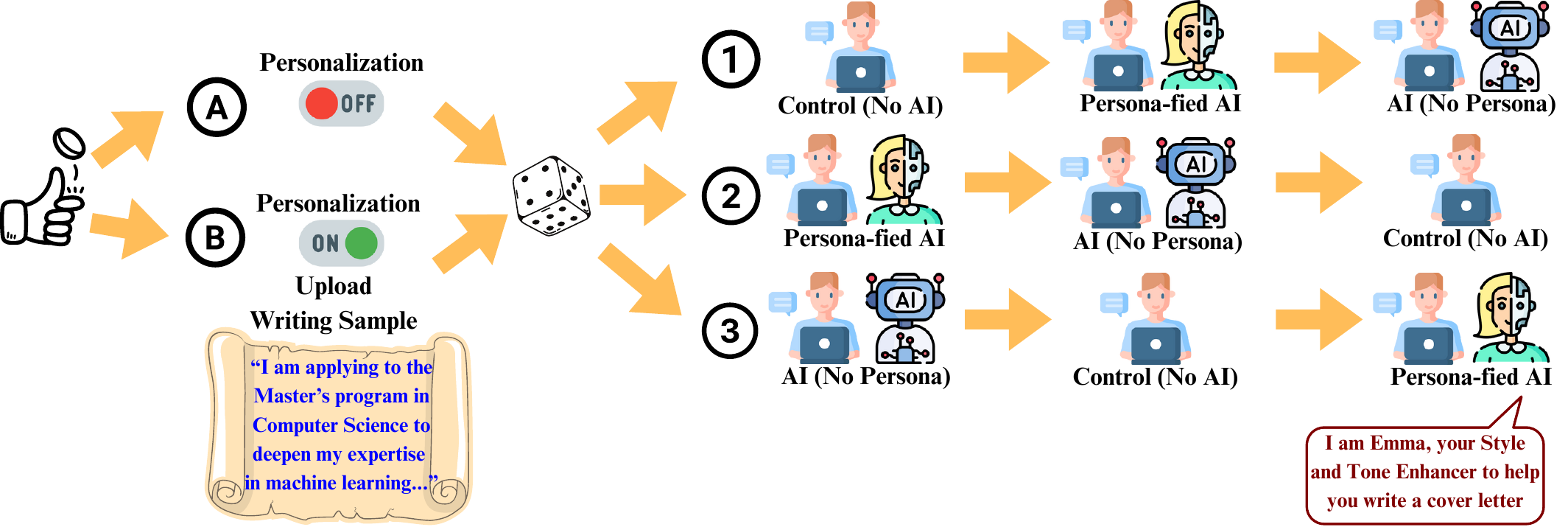}
  \caption{Evaluation design overview. The study uses a mixed design: participants are first randomly assigned to a personalized or non-personalized group (between subjects). Within each group, participants complete three genre-specific writing scenarios (within subjects): an unassisted email baseline (no AI), a generic suggestion mode for proposals (contextual next-sentence suggestions without persona framing), and a persona-framed coaching mode for cover letters (``Emma'' as a style-and-tone coach). Scenario order is counterbalanced using a Latin square. In the personalized group, suggestions are additionally conditioned on a brief writing sample provided by the participant.}
  \label{fig:style-interface}
   \Description{A flow diagram illustrating the experimental design's structure. The diagram splits into two main branches at the top, representing the initial randomization into "Personalized" and "Non-personalized" groups. Under each branch, three sequential boxes show the writing tasks that all participants complete: Control (no AI), AI without persona, and AI with ``Emma'' persona. Arrows indicate that the task order is counterbalanced. For the personalized branch, an additional element shows that participant writing samples inform the AI suggestions. The diagram uses a hierarchical layout to clearly show the between-subjects (personalization) and within-subjects (AI conditions) aspects of the mixed design.}
\end{figure}

Figure~\ref{fig:style-interface} provides an overview of the mixed design, showing the between-subject personalization assignment and the Latin-square counterbalancing of scenario order.

\subsection{Participants}
\label{sec:participants}
We recruited participants from Prolific in Fall 2025. Eligibility targeted adults (18--70) with at least an undergraduate degree residing in English-speaking countries (United States, United Kingdom, Canada) to ensure baseline proficiency for professional writing prompts. To promote data quality, we required $\geq$100 prior Prolific submissions and a $\geq$98\% approval rate. Of 189 starters, we excluded incomplete sessions and unrealistically fast completions ($<$10 minutes), yielding $N{=}176$ valid participants. Compensation included an \$8 base payment (approximately 30 minutes) plus a quality-linked bonus (details in Appendix~\ref{app:front}).

\subsection{Measures}
\label{sec:measures}
We collected self-report outcomes, output quality ratings, and behavioral process measures from interaction logs.

\textbf{Psychological ownership.} Our primary outcome is psychological ownership: the felt sense that the text is ``mine,'' distinct from legal authorship \citep{pierce2001toward}. Following prior work on AI-assisted writing \citep{joshi2025writing}, we compute the mean of four 7-point items capturing personal ownership, responsibility, personal connection, and emotional connection. Full items appear in Appendix~\ref{app:post-survey2}.

\textbf{Cognitive load.} To capture perceived effort, participants completed a Raw NASA-TLX after each task \citep{HART1988139}. We report the overall index and examine composition-relevant subscales (mental demand, effort, frustration), while retaining the full set for completeness (Appendix~\ref{app:post-survey}).

\textbf{Text quality.} Each submission was rated on a 1--7 rubric adapted from \citet{attali2006automated}, covering sentence structure, vocabulary sophistication, coherence, and argumentation. Two trained human raters and one model-based rater (GPT-4o prompted with the same rubric) scored each text independently; we average the three scores to reduce idiosyncratic rater noise. Human-only results and reliability details are reported in the appendix for transparency.

\textbf{Behavioral process measures.} The editor logs interaction telemetry during writing (Section~\ref{sec:system-implementation}). From these logs, we derive: (i) \emph{edit frequency} (keystroke-derived insertions, deletions, replacements), (ii) \emph{AI content incorporation} (estimated fraction of final-draft words originating from accepted suggestions), (iii) \emph{text length} (final word count; all tasks require $\geq$200 words), and (iv) \emph{task duration} (elapsed time from task start to submission). These measures contextualize ownership outcomes by capturing reliance on AI assistance and changes in writing behavior.

\subsection{Analysis approach}
\label{sec:analysis}
We estimate condition differences using OLS regression with participant-clustered, heteroskedasticity-robust standard errors to account for repeated observations. Models include indicator variables for the writing scenarios/assistance modes and controls for task order via Latin-square position. To maintain conservative inference across related outcomes, we control the false discovery rate (FDR) at $q{=}0.05$ using the Benjamini--Hochberg procedure \citep{benjamini1995controlling}. Because assistance modes are intentionally paired with genres, we report scenario-level contrasts in the main text and provide full model specifications, coefficient tables, and robustness checks in Appendix~F.

We treat mediation analyses as exploratory and report them only in the appendix. These analyses examine whether changes in psychological ownership are consistent with a potential pathway linking our interventions to quality ratings. All exploratory analyses and full results are provided in Appendix~F for transparency.

\section{Findings}
\label{sec:findings}

This section reports how the two design levers—persona framing and style personalization—shaped psychological ownership and the writing experience. We focus on findings that directly inform interface-level design decisions; full model outputs and additional process visualizations appear in the appendix.

\subsection{Psychological ownership decreases even as AI assistance remains helpful}

\label{sec:findings-ownership}
Our primary outcome is psychological ownership: whether writers feel the resulting text is ``mine.'' Table~\ref{tab:summary_effects} summarizes the key point estimates. Relative to the unassisted email baseline, both assisted scenarios reduced ownership. In the generic suggestion configuration for proposals, ownership dropped by roughly 1.0 point on a 1--7 scale ($p{<}.001$). In the persona-framed coaching configuration for cover letters, ownership dropped by roughly 0.85 points ($p{<}.001$). In other words, providing fluent, low-friction suggestions reliably weakened felt authorship even when users initiated requests and remained responsible for the final draft.

The persona-framed coaching configuration did not eliminate this decline. Although the ownership drop was slightly smaller than in the generic suggestion configuration, the difference is not large enough to support a claim that persona framing alone preserves authorship. This is consistent with the broader view that role-based framing can shape engagement and interpretation of AI output, but does not automatically transfer ownership when the system contributes substantial language that is easy to adopt.

Style personalization partially restored ownership. Across assisted scenarios, conditioning suggestions on a writer’s prior sample increased ownership by about $+0.43$ points ($p{<}.01$; Appendix~F). Personalization did not fully close the gap relative to the unassisted baseline, but it consistently narrowed it. This pattern supports a ``voice fit'' account: when suggestions better match a writer’s habitual phrasing and register, they feel less like external insertions and more like extensions of the writer’s intent.

We also observe evidence that the personalization benefit is larger in the persona-framed coaching configuration than in the generic suggestion configuration (interaction model in Appendix~F). Interpreted in design terms, this suggests complementary roles: coaching-style framing can scaffold audience and tone, while personalization anchors output to the writer’s voice, and the combination is particularly useful in audience-facing writing such as cover letters. Among the assisted scenarios we studied, the most favorable ownership outcomes occur when coaching-style framing is paired with style conditioning rather than used without voice anchoring.

These patterns appear robust across users. We did not find meaningful differences in ownership effects by prior experience with assistive technologies; the moderation analysis is reported in Appendix~F. The main takeaway for design is therefore not a niche subgroup effect: in our setting, ownership loss is a general consequence of suggestion-based co-writing, and lightweight style personalization offers a consistent partial mitigation.

\subsection{The ownership--efficiency paradox}
\label{sec:findings-paradox}
The ownership results sit alongside a contrasting pattern in effort and quality. Across assisted scenarios, participants reported lower cognitive load while producing text of broadly similar judged quality. AI assistance reduced Raw NASA-TLX scores by approximately $0.88$--$0.91$ points on the 1--7 scale ($p{<}.001$), and personalization did not meaningfully change cognitive load. Quality scores were broadly similar across configurations, though we observe a modest decrease in the generic suggestion configuration (approximately $-0.28$ on a 7-point scale; Table~\ref{tab:summary_effects} and Appendix~F). In other words, the assistance modes we studied reliably reduced perceived effort, while output quality was largely comparable with a small configuration-specific deviation.

Taken together, these findings surface an ownership--efficiency paradox: writers can offload cognitive effort and produce competent text, yet feel less ownership of what they produce. From a design perspective, the paradox matters because it reframes ``helpfulness'' as an incomplete success criterion. A tool can reduce effort without delivering a better authorial experience, and even modest shifts in phrasing and stance can be experienced as a loss of voice. In our data, the most direct lever for reducing this experiential cost is not persona framing in isolation, but improving stylistic fit through personalization, which increases ownership and adoption without materially changing workload.

To make this tradeoff legible, Table~\ref{tab:summary_effects} summarizes point estimates for the outcomes most relevant to design. Rather than presenting a dense set of process plots in the main body, we highlight the core pattern: ownership decreases under assistance, workload decreases, adoption increases, and personalization selectively restores ownership while further increasing adoption. Additional process regressions (e.g., edits, time, word count) are reported in Appendix~F.

\begin{table}[t]
\centering
\small
\caption{Summary of key effects (point estimates). ``Generic'' and ``Persona'' report scenario coefficients relative to the unassisted email baseline. ``Personalized'' reports the between-subject main effect of style personalization. Stars reflect FDR-adjusted inference. Full tables appear in the appendix.}
\label{tab:summary_effects}
\begin{tabular}{lccc}
\toprule
\textbf{Outcome} & \textbf{Generic vs. baseline} & \textbf{Persona vs. baseline} & \textbf{Personalized vs. not} \\
\midrule
Ownership & $-1.03^{***}$ & $-0.85^{***}$ & $+0.43^{**}$ \\
Cognitive load (NASA-TLX) & $-0.88^{***}$ & $-0.91^{***}$ & $+0.05$ \\
Quality & $-0.28^{**}$ & $-0.18$ & $-0.01$ \\
AI content incorporation & $+0.31^{***}$ & $+0.28^{***}$ & $+0.05^{*}$ \\
\bottomrule
\multicolumn{4}{p{0.98\linewidth}}{\footnotesize Notes: ``Generic'' corresponds to the generic suggestion mode (proposal) and ``Persona'' to the persona-framed coaching mode (cover letter). AI content incorporation is a fraction (e.g., $0.31 \approx 31\%$). $^{*}p{<}.05,\ ^{**}p{<}.01,\ ^{***}p{<}.001$ (FDR-adjusted).}
\end{tabular}
\end{table}

\subsection{Adoption patterns}
\label{sec:findings-adoption}
Behavioral logs provide a complementary view of how writers used the AI assistance. In both AI assisted scenarios, participants incorporated a substantial fraction of AI-generated text into their final drafts—approximately 28--30\% on average ($p{<}.001$ relative to baseline). This level of adoption is consistent with the interaction model: suggestions were short, low-friction to accept, and designed to be easy to integrate into the draft.

Style personalization increased adoption by about five percentage points ($p{<}.05$). Importantly, this increase coincided with higher psychological ownership rather than lower ownership. In design terms, higher adoption with higher ownership is informative: when suggestions are aligned to the writer’s voice, accepting them appears less like ceding authorship and more like extending one’s own expression with low-effort phrasing help. This supports a practical interpretation of personalization as \emph{voice anchoring} rather than automation: it can increase the utility of suggestions while reducing their experiential cost.

AI assistance also changed revision behavior. Relative to the unassisted baseline, writers made fewer edits under both assisted scenarios ($p{<}.001$). Personalization did not significantly change edit counts ($p{>}0.05$). Together with stable task duration in the appendix, this suggests that writers did not simply finish faster; instead, they revised less while incorporating more suggested text, consistent with shifting effort from sentence-level wording toward higher-level content decisions.

\subsection{Qualitative Themes: voice fit, rhetorical defaults, and where the tool shows through}
\label{sec:findings-qual}
To clarify how personalization changes the experience of authorship, we examined drafts and accepted suggestions across conditions. Beyond the high-level ``voice fit versus generic phrasing'' contrast, three qualitative patterns were recurrent: (i) personalization often preserved a writer’s \emph{register} and rhetorical cadence, (ii) non-personalized assistance often converged to a narrow set of \emph{professional defaults} (enthusiasm, alignment, impact), and (iii) persona framing tended to stabilize \emph{stance} (coach-like encouragement) without reliably stabilizing \emph{voice} \citep{lee2024design,zhao2025making}.

\paragraph{\underline{Theme 1: Voice fit through register and cadence}}
In personalized conditions, suggestions frequently mirrored the writer’s register—formal versus conversational—and matched cadence (e.g., clause length, use of intensifiers, and preferred metaphors). This made suggestions easier to incorporate without rewriting.

\begin{quote}
\textbf{Personalized + persona-framed coaching (voice-aligned phrasing).} When seeded with a graduate-school personal statement, the assistant adapted to formal, aspirational prose, offering phrases such as ``it resonates with my passion for pushing boundaries'' and ``I thrive in dynamic environments where innovation meets impact.'' The suggestions matched the writer’s established register and could be adopted with minimal stylistic friction.
\end{quote}

A second recurring instance of fit was that personalization could preserve a writer’s habitual \emph{structure}—for example, using parallel constructions, hedges, or reflective framing—rather than defaulting to generic corporate boilerplate. Even when the informational content was conventional (e.g., enthusiasm, qualifications), the phrasing felt more plausibly authored by the participant.

\paragraph{\underline{Theme 2: Non-personalized ``professional rhetoric'' as a default}}
Without style conditioning, suggestions frequently converged on broadly applicable claims that sound appropriate in many contexts but are weakly tied to any specific authorial voice. In cover letters, this often surfaced as stock phrases about ``leveraging expertise,'' ``innovative mindset,'' and ``aligning with the company’s vision.'' This convergence toward generic rhetorical defaults is consistent with broader concerns that language-model assistance can reduce diversity and increase stylistic homogenization in writing \citep{padmakumar2023does,li2024value}.

\begin{quote}
\textbf{Non-personalized + persona-framed coaching (generic professional rhetoric).} Without style conditioning, suggestions often fell back on broadly applicable claims such as ``leverage my technical expertise and innovative mindset'' and ``aligns perfectly with InnovateTech's vision.'' The language maintained a consistent tone but was not anchored to any individual voice, making it read more like ``from the tool'' than ``from me.''
\end{quote}

In several drafts, this default rhetoric also introduced a subtle mismatch: it would amplify confidence and superlatives beyond what the participant’s surrounding text supported. Writers could still accept these suggestions because they were fluent and locally coherent, but the added intensity made the insertion feel externally authored.

\paragraph{\underline{Theme 3: Persona stabilizes stance more than voice}}
Persona framing (``Emma'' as a coach) tended to create a consistent stance—encouraging, polished, and audience-aware—across writers. This helped maintain cover-letter conventions (e.g., enthusiasm, fit, closing professionalism). However, unless paired with personalization, the persona’s stance sometimes dominated the local writing voice. This aligns with prior observations that social framing can shape how writers interpret AI support while leaving open questions about authorship and ownership \citep{gero2023social}. It is also consistent with our quantitative interaction: the persona-framed coaching scenario benefited most from personalization, suggesting that personas may scaffold audience and tone but still require voice anchoring to preserve authorship.

Overall, the qualitative patterns help interpret the quantitative results. Personalization did not primarily improve judged quality or reduce effort; instead, it changed how suggestions \emph{fit} into a writer’s draft. When suggestions echoed the writer’s register and cadence, they were easier to adopt (higher incorporation) and felt more like extensions of the writer’s intent (higher ownership). When suggestions defaulted to generic professional rhetoric, they were still usable, but they contributed to the sense that the draft reflected the assistant’s voice rather than the writer’s. Extended examples across conditions appear in Appendix~G, and full regression outputs appear in Appendix~F.

\section{Design Patterns for Preserving Authorship}
\label{sec:patterns}

Our findings support a set of design patterns for AI writing assistance that preserves authorship while maintaining efficiency benefits. Each pattern addresses a specific threat to psychological ownership and is grounded in our evaluation evidence. Taken together, they offer a practical vocabulary for designing professional co-writing tools that reduce effort without treating authorship as collateral damage.

\paragraph{How to read these patterns.}
We did not factorially isolate every interaction choice in the interface. Instead, we distill patterns from the system we built and evaluated, supported by observed outcomes and existing literature on ownership and agency. Each pattern synthesizes (i) artifact design decisions, (ii) evaluation findings, and (iii) prior work on human--AI collaboration. They should be understood as design knowledge—actionable guidance for practitioners—rather than isolated causal claims about individual UI elements. We present five patterns implemented in our system, followed by design implications for future work.

\subsection{Pattern 1: On-demand initiation preserves agency}
\textbf{Problem.} Proactive suggestions can interrupt flow and make writers feel managed rather than supported. Unsolicited assistance shifts the locus of control from writer to system, increasing the chance that AI participation feels imposed rather than invited.

\textbf{When it applies.} Any AI writing assistance context where user agency matters, especially tasks that require concentration, planning, or creative flow. This is particularly relevant in professional writing, where users may want assistance at specific moments (e.g., phrasing a sentence, adjusting tone) but not continuous intervention.

\textbf{Pattern.} Let users explicitly request suggestions rather than pushing them automatically. The writer decides when to invite the system into the writing process.

\textbf{Implementation.} Provide a clear, one-action entry point for assistance (e.g., \textit{Ask AI} / \textit{Ask Emma}) and generate suggestions only on request. Avoid automatic completions, pop-ups, or timed interruptions. Preserve a predictable loop: request $\rightarrow$ preview $\rightarrow$ accept/ignore.

\textbf{Evidence.} In our evaluation, suggestions entered the draft only through explicit user requests, and writers controlled when to solicit help. While ownership decreased relative to unassisted writing, it was not eliminated, consistent with the view that initiation control can preserve some sense of agency even when suggestions are easy to adopt. This interpretation aligns with prior work emphasizing adjustable autonomy and user control as foundations for effective human--AI collaboration \citep{amershi2019guidelines,feigh2014requirements}.

\textbf{Tradeoffs and boundary conditions.} On-demand designs can reduce discovery of helpful suggestions and require users to remember that help is available. Some contexts (e.g., accessibility support, heavy boilerplate) may benefit from optional proactive assistance; when used, proactive features should be clearly controllable and easy to disable.

\subsection{Pattern 2: Micro-suggestions over takeovers}
\textbf{Problem.} Long AI completions can overwrite voice and reduce felt authorship even when output quality is maintained. Extended generations increase the risk that the system becomes the de facto author, with the user shifting from writing to approving.

\textbf{When it applies.} Sentence- and paragraph-level assistance in genres where individual voice matters and accountability is tied to the final text (e.g., cover letters, proposals, client-facing communication).

\textbf{Pattern.} Bound suggestion length to short, adoptable increments. Each suggestion should feel like a contribution to the writer’s draft, not a replacement for it.

\textbf{Implementation.} Cap suggestions to a short span (e.g., one sentence) and require incremental acceptance (e.g., accept up to punctuation). Make acceptance lightweight (e.g., Tab or a single button) while preserving easy revision after acceptance. Use conservative defaults that discourage multi-paragraph completions unless explicitly requested.

\textbf{Evidence.} Writers incorporated a substantial fraction of AI text (approximately 28--30\%) while judged quality remained stable. Ownership was reduced, but not eliminated, suggesting that bounded suggestions limited how much any single acceptance could displace voice. Qualitative examples show micro-suggestions that extend local phrasing rather than replacing entire passages, especially when paired with personalization (Section~\ref{sec:findings-qual}).

\textbf{Tradeoffs and boundary conditions.} Micro-suggestions may feel insufficient for users who want substantial drafting help and can require more frequent requests during long-form writing. For boilerplate-heavy tasks, longer completions may be appropriate, but should be explicitly initiated and accompanied by stronger provenance support.

\subsection{Pattern 3: Voice anchoring via style personalization}
\textbf{Problem.} Generic suggestions can read as ``from the tool'' rather than ``from me,'' reducing psychological ownership even when the content is useful. When suggestions mismatch the writer’s habitual register or cadence, adoption can feel like importing someone else’s voice.

\textbf{When it applies.} Any context where maintaining individual voice matters—professional communication, creative writing, and personal correspondence—especially when writing is tied to identity and self-presentation.

\textbf{Pattern.} Derive a lightweight style profile from a writer’s prior text and condition suggestions on it. Align AI output with habitual patterns of expression.

\textbf{Implementation.} Ask users for a brief writing sample (or use prior opt-in artifacts), extract a small set of interpretable style traits (e.g., tone, formality, sentence structure, lexical choice, rhetorical tendencies), and inject these constraints into suggestion prompts. Keep the profile lightweight enough that users do not need extensive configuration to benefit.

\textbf{Evidence.} Style personalization increased ownership by approximately $+0.43$ points and increased AI text incorporation by about $+5$ percentage points. The combination—higher adoption with higher ownership—suggests that voice-aligned suggestions can be experienced as extensions of the writer’s intent rather than as automation. This pattern is consistent with self-extension accounts in which artifacts that align with identity feel more ``mine'' \citep{belk1988possessions}. Qualitative examples show personalization preserving register and cadence, whereas non-personalized suggestions often defaulted to generic professional rhetoric (Section~\ref{sec:findings-qual}).

\textbf{Tradeoffs and boundary conditions.} Personalization requires collecting writing samples and thus raises privacy and consent concerns. Style extraction can be imperfect or overly rigid, and voice is often context-dependent (e.g., an email versus a cover letter). In our data, the ownership benefit of personalization was strongest when paired with persona-framed coaching, suggesting that voice anchoring is particularly valuable when the interface also provides an audience scaffold.

\subsection{Pattern 4: Persona as audience scaffold, not author}
\textbf{Problem.} AI assistants can feel like ghostwriters, making the resulting text feel authored by the machine rather than the human. When persona framing is interpreted as ``the assistant writing,'' it can intensify authorship displacement even when the output is polished.

\textbf{When it applies.} Audience-facing genres where tone, stance, and reader expectations matter—cover letters, professional correspondence, outreach, and client communication.

\textbf{Pattern.} Frame the AI as a role-based coach (e.g., ``style and tone enhancer'') rather than a co-author. Use persona primarily to scaffold audience awareness and appropriate register, not to transfer authorship.

\textbf{Implementation.} Present the persona with an explicit supportive role (coach, editor, tone advisor) and describe the assistant as helping the writer refine \emph{their} draft rather than producing content on their behalf. Ensure that persona prompts emphasize edits, refinements, and audience fit. Keep suggestions bounded and adoptable so the writer remains the driver of composition.

\textbf{Evidence.} Persona framing alone did not restore ownership: the persona-framed coaching scenario showed a substantial ownership drop relative to baseline. However, the persona $\times$ personalization interaction indicates that persona scaffolding worked best when paired with voice anchoring. This supports a practical reading: personas can help with audience and tone, but they do not by themselves make the text feel authored by the writer; voice alignment via personalization is critical for ownership.

\textbf{Tradeoffs and boundary conditions.} Persona framing adds interface and conceptual complexity, and effectiveness depends on role--genre fit. A persona that feels mismatched to the task can backfire. Designers should treat persona choices as genre-sensitive defaults rather than universally beneficial features.

\subsection{Pattern 5: Provenance cues at the point of decision}
\textbf{Problem.} Writers may accept suggestions without fully registering what they are adopting, leading to text that later feels alien or externally authored. When the boundary between writer text and AI text becomes cognitively invisible, ownership and accountability can suffer.

\textbf{When it applies.} Any AI writing context where authorship, accountability, or disclosure matters, especially in professional settings where writers are responsible for the final content.

\textbf{Pattern.} Visually distinguish AI suggestions from user text during the review and acceptance process, and make adoption explicit. The moment of adoption should feel like a deliberate choice.

\textbf{Implementation.} Display suggestions in a visually distinct form (e.g., light gray) prior to adoption. Require an explicit acceptance action (e.g., Tab or an \textit{Accept Suggestion} button), after which the text becomes part of the editable draft. Avoid silent insertion of AI text without a clear accept/reject step.

\textbf{Evidence.} In our system, AI text entered the draft only through explicit user choice: writers requested suggestions, previewed them in gray, and accepted them through a deliberate action. This supported provenance legibility during writing and enabled precise provenance tracking in logs (e.g., estimating AI content incorporation). While this does not eliminate ownership loss, it provides a concrete control point where writers can reflect on whether a suggestion matches their intent before adopting it.

\textbf{Tradeoffs and boundary conditions.} Provenance is legible during writing but not persistent in the final document: once accepted, AI-originated text is visually indistinguishable from human-written text. Persistent provenance mechanisms are therefore design implications for future tools rather than features we implemented.

\section{Design Implications and Future Directions}
\label{sec:implications}

Beyond the implemented patterns, our findings suggest additional design directions that merit future exploration. These implications extend provenance and control mechanisms that our current system does not fully realize.

\paragraph{\underline{Implication 1: Persistent provenance and AI contribution indicators}}
Pattern~5 provides provenance cues at the point of decision, but the final document contains no persistent record of what originated from AI. Future systems could implement a live ``AI \% of draft'' meter showing cumulative AI contribution, subtle but persistent visual marking of AI-originated passages (e.g., low-saturation highlighting or margin glyphs), and a revision history that distinguishes human edits from AI acceptances. These mechanisms would extend provenance awareness beyond the writing session into the final artifact, supporting accountability and enabling writers to revisit and rework adopted spans when ownership or disclosure concerns arise.

\paragraph{\underline{Implication 2: ``Hold My Voice'' control}}
Our findings suggest that personalization primarily helps by anchoring voice. A user-adjustable control could increase style fidelity when ownership matters most. Writers could toggle between an ``explore alternatives'' mode (looser style matching that prioritizes novelty) and a ``hold my voice'' mode (stricter alignment with the writer’s profile) depending on task demands and personal preference. Such controls would make stylistic influence tunable rather than implicit.

\paragraph{\underline{Implication 3: ``Rephrase in My Voice'' action}}
Writers may want to adopt an idea while rejecting its phrasing. A one-click ``rephrase in my voice'' action could restyle an accepted suggestion using the writer’s profile, addressing cases where the content is useful but the voice is mismatched. This supports adoption while maintaining stylistic ownership and reduces the need for manual rewriting after accepting a suggestion.

These implications extend our patterns toward more controllable, transparent AI writing assistance. Realizing them requires attending to design details—where controls appear, how provenance is visualized, and what ``voice'' means computationally—that shape whether AI assistance supports or supplants authorship.

\section{Limitations and Ethical Considerations}
\label{sec:limitations-ethics}

\subsection{Limitations}
Our findings should be interpreted with several scope conditions in mind. First, our evaluation uses a \emph{matched-task} design in which each assistance mode is paired with a specific professional genre (unassisted email baseline, generic suggestions for proposals, and persona-framed coaching for cover letters). This reflects our position that AI writing assistance should be genre-appropriate rather than one-size-fits-all, and it improves ecological realism. However, it also limits causal claims about assistance modes independent of genre. Accordingly, the patterns we distill should be understood as emerging from \emph{genre--mode configurations} rather than as isolated effects of persona framing or suggestion style.

Second, several interaction choices emphasized in our design patterns were held constant rather than experimentally varied. In particular, on-demand initiation, micro-suggestions, and point-of-decision provenance cues (Patterns 1, 2, and 5) were implemented across assisted scenarios. Evidence for these patterns therefore comes from (i) observing that ownership was reduced but not eliminated despite substantial AI contribution, (ii) behavioral traces indicating deliberate request-and-accept interactions, and (iii) support from prior literature on agency and control, rather than from factorial manipulation of those UI elements.

Third, our study is a short-session online experiment. Single-session writing tasks may not capture how ownership dynamics evolve over longer projects, revision cycles, or repeated tool use, where writers may either acclimate to assistance or become more sensitive to voice drift. Fourth, we tested one model (GPT-4o) and one persona instantiation (``Emma''); other models, persona roles, or prompting strategies may yield different patterns. Fifth, personalization was derived from a single writing sample; richer profiles (multiple samples, explicit preference settings, or adaptive learning) might strengthen or complicate the personalization effect. Finally, our participant pool was drawn from English-speaking countries and required at least some college education, which supports baseline proficiency for professional prompts but limits generalization to other populations, languages, and workplace contexts.

\subsection{Ethical Considerations}
AI-assisted writing raises questions about authorship, attribution, and integrity in professional and academic contexts. While our patterns support transparency at the point of adoption, they do not resolve normative questions about when AI use should be disclosed or how credit should be assigned. Systems that make AI assistance easier to adopt also risk shifting writing norms in ways that are not always visible to users.

Personalization introduces additional responsibilities around privacy and consent because it requires collecting writing samples. Practical safeguards include explicit opt-in, transparent retention policies, minimum necessary data collection, and user-controlled deletion of samples and derived profiles. Designers should treat style profiles as sensitive user data and avoid repurposing them beyond the immediate personalization function without clear consent.

Finally, there are broader ecosystem risks. If AI suggestions converge toward generic ``professional'' language, widespread adoption could reduce diversity of expression and encourage stylistic homogenization; personalization may counteract this tendency, but the tension merits ongoing attention. Reduced cognitive load is a clear benefit, yet long-term reliance on assistance could also affect writing skill development. Interfaces should therefore support learning and voice development—not just task completion—by keeping AI influence transparent, tunable, and reversible.

\section{Conclusion}
\label{sec:conclusion}

AI writing assistance can reduce cognitive burden and maintain output quality while eroding writers' felt authorship—an ownership--efficiency paradox that poses a design challenge. Our work addresses this challenge through an ownership-aware co-writing editor and an evaluation that surfaces how design choices shape the writing experience in professional genres.

We contribute five design patterns grounded in our findings: on-demand initiation preserves agency; micro-suggestions prevent takeover; style personalization anchors voice; persona framing scaffolds audience without transferring authorship; and provenance cues at the point of decision support deliberate adoption. These patterns synthesize artifact design, evaluation evidence, and prior literature into actionable guidance for practitioners building AI writing tools.

More broadly, our results suggest that AI writing assistance is not only a productivity feature but also an identity-facing system. The text people produce with AI help becomes part of how they present themselves professionally and personally. Designing for authorship therefore means attending to this entanglement—building systems whose influence is transparent, tunable, and ultimately in service of the human voice they support.

\bibliographystyle{ACM-Reference-Format}
\bibliography{sample-base}

\appendix
\section{Experiment Prompts}\label{app:prompts}
\subsection{Prompts to Build persona-framed coaching Assistant `Emma'}\label{app:style-prompt}
\begin{tcolorbox}[colback=blue!5!white, colframe=blue!75!black, breakable]
You are Emma, an AI Style and Tone Enhancer assistant. Your role is to help a job applicant write a cover letter for a mid-level position at InnovateTech Solutions, a fast-growing tech company known for its innovative software products. The job description emphasizes the need for strong technical skills, creativity in problem-solving, and excellent communication abilities. The applicant has relevant experience at a smaller tech firm called InfoSoft Solutions. You should provide text continuations that improve the letter's writing style, tone, and overall effectiveness.\\ When the user requests assistance, analyze the current text and generate a continuation that:
\begin{enumerate}
    \item Maintains a balance between professionalism and personality
    \item Uses enhanced vocabulary and phrasing for clarity and impact
    \item Improves the structure and flow of the cover letter
    \item Addresses any key points not yet covered (introduction, skills, company fit, enthusiasm)
\end{enumerate}
\indent Your generated text should:
\begin{enumerate}
    \item Seamlessly continue from where the user left off
    \item Add EXACTLY one sentence with LESS than 20 words that improves the overall quality and effectiveness of the cover letter.    
\end{enumerate}
\indent Remember that the cover letter should introduce the applicant, highlight relevant skills and experiences, demonstrate an understanding of the company's culture, explain why they're a great fit, and convey a unique voice while maintaining a professional tone. Do NOT preface your suggestions with any explanations. Simply provide the continuation as if it were part of the original text. Here is what has been written:
\end{tcolorbox}
\subsection{Prompts for AI (no-persona) Assistance}\label{app:perspective-prompt}
\begin{tcolorbox}[colback=blue!5!white, colframe=blue!75!black, breakable]
Generate suggestions to complete the sentence or provide the next sentence based on the text that has been written so far.\\
Your generated text should:
\begin{enumerate}
    \item Seamlessly continue from where the user left off
    \item Add EXACTLY one sentence with LESS than 20 words
\end{enumerate}
Do NOT preface your suggestions with any explanations. Simply provide the continuation as if it were part of the original text. Here is what has been written:
\end{tcolorbox}

\subsection{Prompt to Summarize the Writing Styles of Participants} \label{app:summarize}
\begin{tcolorbox}[colback=blue!5!white, colframe=blue!75!black, breakable]
Analyze the following text and directly list 5 specific unique aspects of the author's writing style, focusing on tone, voice, sentence structure, word choice, use of rhetorical devices, narrative style, and distinctive grammar or syntax patterns. Your output should directly start with the five bullet points without any preface. Here is the text:
\end{tcolorbox}

\subsection{Prompts for Personalization}
For personalization, the following prompt will be appended to the prompts \ref{app:style-prompt} or \ref{app:perspective-prompt}.
\begin{tcolorbox}[colback=blue!5!white, colframe=blue!75!black, breakable]
Before generating suggestions, carefully review the following summary of the user's writing style: ... Here is what has been written: \end{tcolorbox}

\section{The Front Page of The Experimental Interface}\label{app:front}
\begin{tcolorbox}[colback=blue!5!white, colframe=blue!75!black, breakable, title=\centering Welcome to
Study on Human-AI Collaborative Writing]
\textbf{Principal Investigator}: [Anonymous for submission]
\\\\
You are invited to participate in a research study on human-AI collaborative writing. We aim to explore how AI can assist and enhance the writing process in various contexts. As a participant, you'll engage in a series of three writing tasks: one without AI assistance and two with different kinds of AI support.
\\\\
Your participation will help us understand how AI impacts writing quality, efficiency, and writer satisfaction across different types of writing tasks. By analyzing these factors, we seek to develop more effective AI writing assistants that complement human creativity and expertise.
\\\\
Join us in shaping the future of AI-assisted writing and contribute to the development of tools that empower human writers in diverse and meaningful ways!
\\\\
If you agree to be part of the research study, you will be asked to read and sign (by clicking the "I consent" button) the consent form. After submitting the consent form, you will receive a pre-survey which will be followed by multiple writing prompts. Each writing task will require you to write on a prompt with or without the help of an AI. After completing each sub-task, you will be asked to complete a short survey. Throughout this task, you should refrain from multi-tasking, i.e., checking your phone, email, or the Internet.
\\\\
\textbf{Compensation}: You will be compensated \$8.00 as a base rate for participating in this study. Based on the quality of your written texts, you can earn additional compensation. If your average writing quality score is higher than 4 but less than 6 out of 7, you will receive an extra \$5.00 (total of \$13.00). If your average writing quality score is 6 or higher out of 7, you will receive an extra \$10.00 (total of \$18.00).
\\\\
Your writing quality will be assessed using various criteria, including the complexity of sentence structure, the use of advanced vocabulary, the coherence of the text, and the effectiveness of arguments presented.
\\\\
\textbf{Writing with the help of an AI}: You will complete three writing tasks: one without AI assistance and two with different types of AI support. For tasks involving AI, you can request suggestions by clicking a button. The AI is designed to assist you in various ways, which will be explained before each relevant task. The specific characteristics of the AI assistance may vary based on your assigned condition. You have the freedom to modify, partially use, or completely reject the AI suggestions, and can always request new ones. After each task, you'll be asked about your experience. This study aims to understand how different types of AI assistance affect the writing process and writers' perceptions of their work.
\\\\
\textbf{Benefits of the research}: While there may not be any immediate personal benefits for participants, their involvement will greatly aid the research team and the wider AI community in understanding the intricacies, advantages, and obstacles of writing collaboratively with AI.
\\\\
\textbf{Risks and discomforts}: You might experience fatigue. You can pause during the task and continue from where you left off. Also, you can stop participating in the study at any time without any cost to you.
\\\\
\textbf{Data security and storage}: We will protect the confidentiality of your research records by not using any identifiable information, such as your name or email address, to identify you on any study records. Instead, you will be assigned a unique random identifier. To minimize the risk of a data breach, we store all the content on a password-protected server behind a firewall minimizing any such breaches.
\\\\
Participating in this study is entirely voluntary. You must be 18 years or older in order to participate in this study.. Even if you decide to participate now, you may change your mind and stop at any time. You may choose to stop participating in the study at any time for any reason. You can request to delete your profile at any time and withdraw from the study.
If you have questions about this research study, please contact [Anonymous for submission]
As part of their review, the Institutional Review Board (IRB) has determined that this study is no more than minimal risk and exempt from on-going IRB oversight.
\\\\
\ding{113} I consent \quad\quad\quad\quad \ding{113} I do not consent
\end{tcolorbox}

\section{Supportive Plots of the Experimental Interface}\label{app:interface}
\begin{figure}[H]
  \centering
  \includegraphics[width=\linewidth]{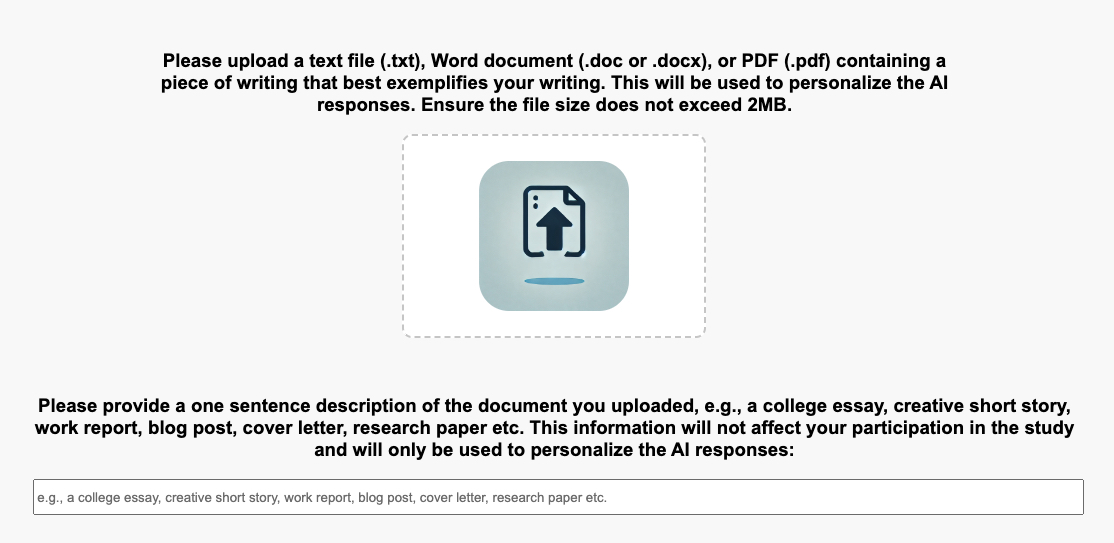}
  \caption{Experimental interface for uploading participants' writing samples.}
  \label{fig:upload}
  \Description{A screenshot showing the experimental interface where participants can upload their writing samples, with input fields and an upload button.}
\end{figure}

\begin{figure}[H]
  \centering
  \includegraphics[width=\linewidth]{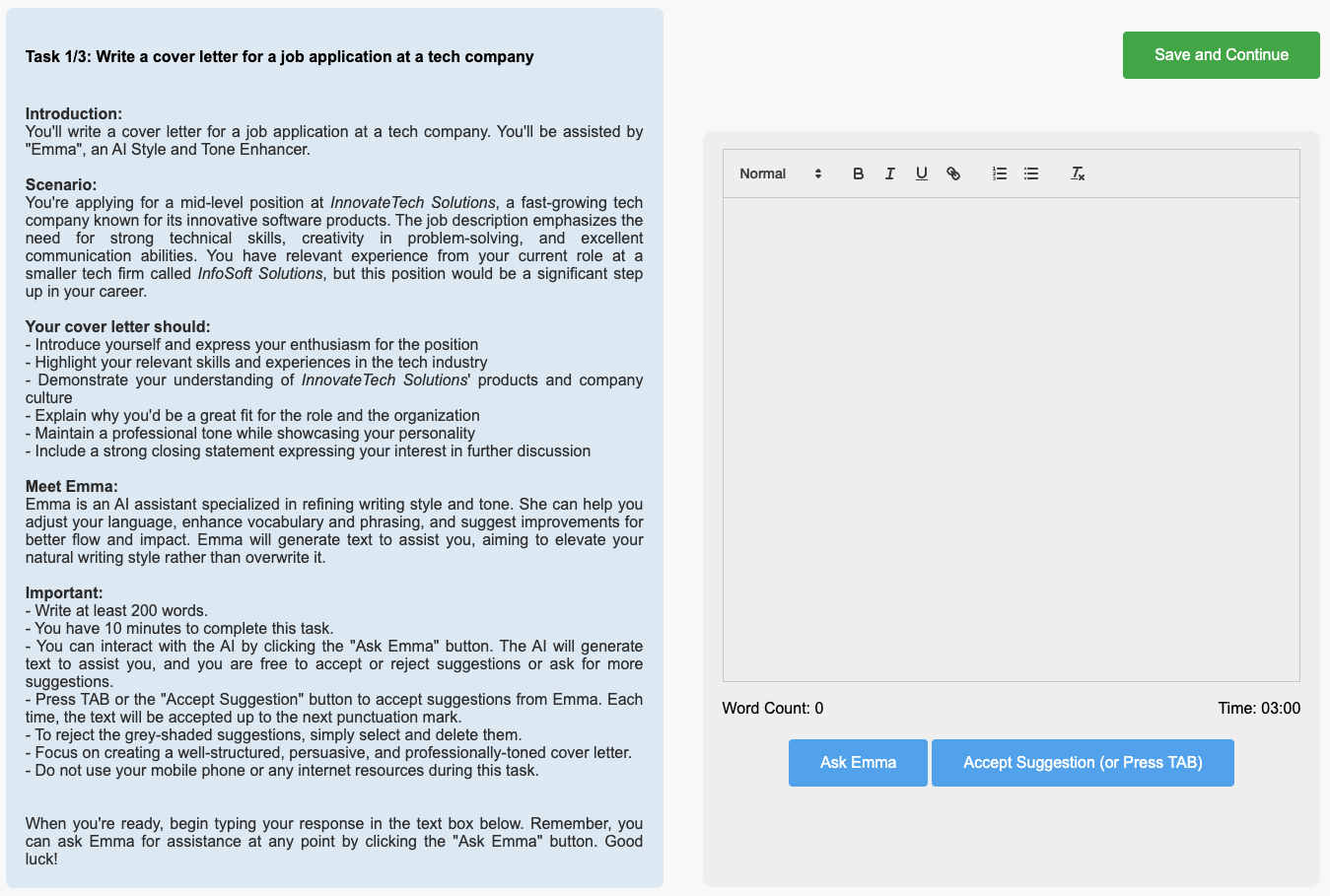}
  \caption{Experimental interface for persona-framed coaching (style-enhancing persona) condition.}
  \label{fig:style}
  \Description{A screenshot showing the experimental interface for the style-enhancing persona condition, featuring a text editor and AI-generated style suggestions.}
\end{figure}

\begin{figure}[H]
  \centering
  \includegraphics[width=\linewidth]{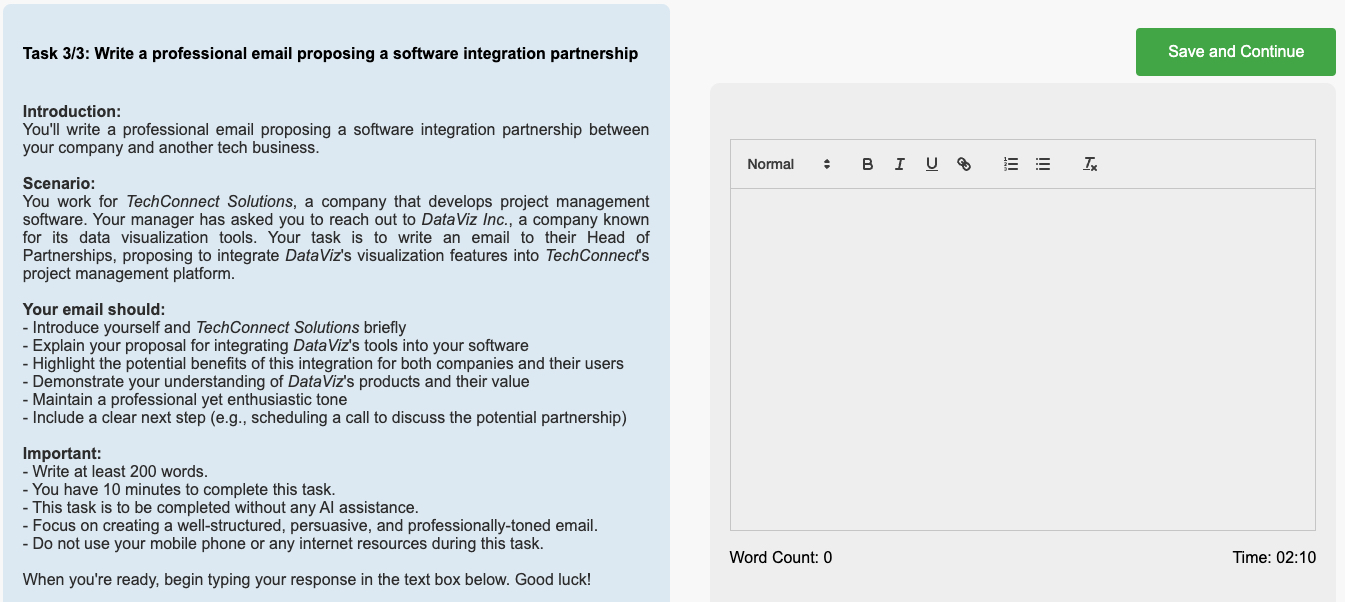}
  \caption{Experimental interface for Control (No AI) condition.}
  \label{fig:control}
  \Description{A screenshot showing the experimental interface for the control condition without AI, including a basic text editor.}
\end{figure}

\begin{figure}[H]
  \centering
  \includegraphics[width=\linewidth]{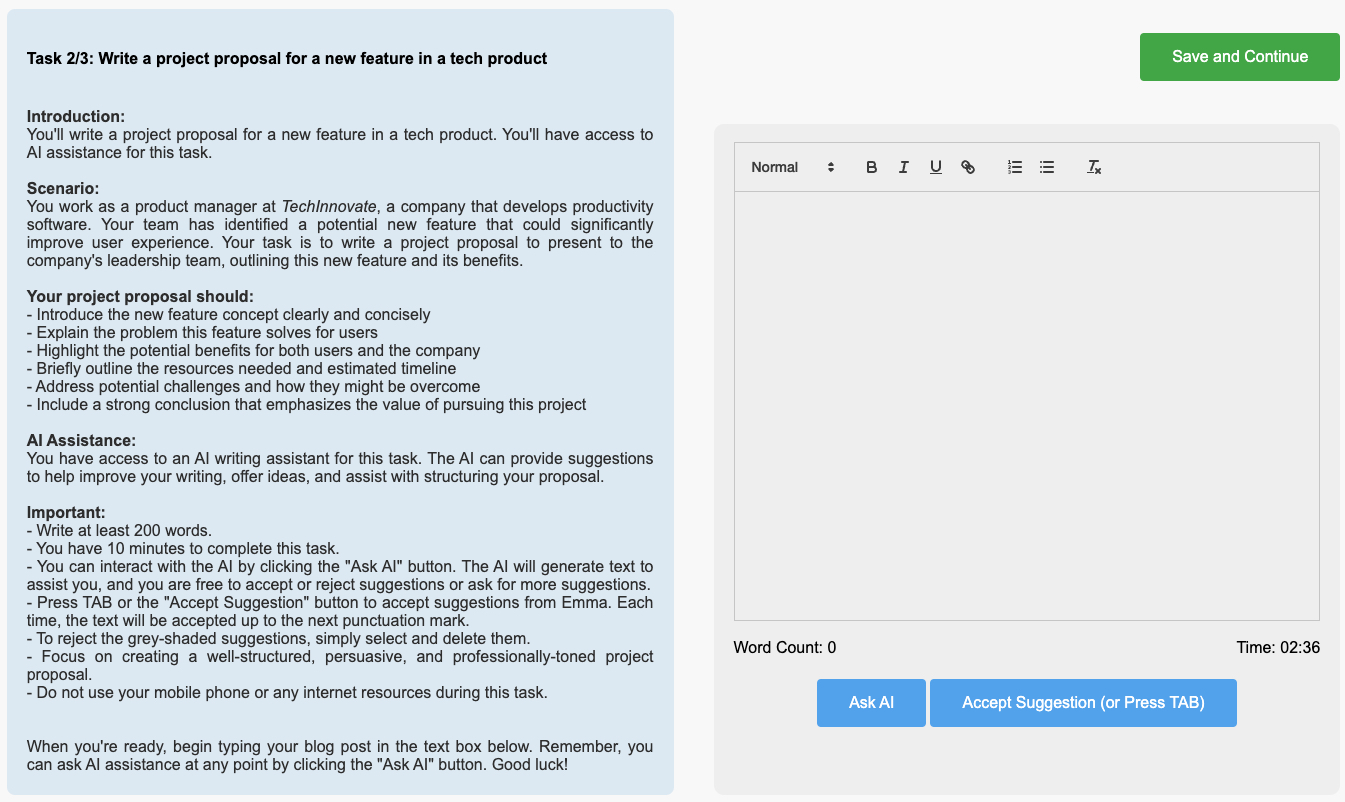}
  \caption{Experimental interface for AI (no-persona) condition.}
  \label{fig:perspective}
  \Description{A screenshot showing the experimental interface for the no-persona AI condition, with a text editor and generic AI suggestions.}
\end{figure}

The interface for uploading writing samples is shown in Figure \ref{fig:upload}. The interfaces of control, AI (no-persona), and persona-framed coaching conditions are shown in Figure \ref{fig:control}, \ref{fig:style}, and \ref{fig:perspective}. Note: The ``10 minutes'' message in the interface was intended as guidance to keep the session moving; it was not enforced as a hard cutoff. Participants could exceed 10 minutes to reach the 200-word minimum and submit.

\section{API Parameters}\label{app:api}
The model is `GPT-4o-2024-08-06'. The maximum number of output tokens is 30. The temperature is 0.8 and other parameters are set as default.

\section{Surveys}
\subsection{Pre-task Survey}\label{app:pre-survey}
\begin{enumerate}[label=\arabic*.]

    \item What is your age range?
    \begin{itemize}
        \item $\bigcirc$ 18 -- 25
        \item $\bigcirc$ 25 -- 35
        \item $\bigcirc$ 35 -- 45
        \item $\bigcirc$ 45 -- 55
        \item $\bigcirc$ > 55
    \end{itemize}

    \item What is your gender?
    \begin{itemize}
        \item $\bigcirc$ Female
        \item $\bigcirc$ Male
        \item $\bigcirc$ Non-binary
        \item $\bigcirc$ I do not wish to specify
    \end{itemize}

    \item What is your level of education?
    \begin{itemize}
        \item $\bigcirc$ High-school diploma
        \item $\bigcirc$ College degree
        \item $\bigcirc$ Graduate degree
    \end{itemize}

    \item What is your first language?
    \begin{itemize}
        \item $\bigcirc$ English
        \item $\bigcirc$ Not English
    \end{itemize}

    \item How do you rate your level of English proficiency?
    \begin{itemize}
        \item $\bigcirc$ I understand English, but I am not fluent
        \item $\bigcirc$ I am fluent
        \item $\bigcirc$ I am a native speaker
    \end{itemize}

    \item What is your level of writing expertise?
    \begin{itemize}
        \item $\bigcirc$ I am a professional writer
        \item $\bigcirc$ I am not a professional writer, but I write regularly
        \item $\bigcirc$ I am not a professional writer, and I do not write regularly
    \end{itemize}

    \item What is your ethnicity?
    \begin{itemize}
        \item $\bigcirc$ Hispanic or Latino
        \item $\bigcirc$ Not Hispanic or Latino
        \item $\bigcirc$ Prefer not to say
    \end{itemize}

    \item What is your race?
    \begin{itemize}
        \item $\bigcirc$ American Indian or Alaska Native
        \item $\bigcirc$ Asian
        \item $\bigcirc$ Black or African American
        \item $\bigcirc$ Native Hawaiian or Other Pacific Islander
        \item $\bigcirc$ White
        \item $\bigcirc$ Prefer not to say
    \end{itemize}

    \item Have you used assistive technology in writing before this study?
    \begin{itemize}
        \item $\bigcirc$ Yes, only spell check, or grammar check, or autocomplete
        \item $\bigcirc$ Yes, more advanced writing assistants
        \item $\bigcirc$ No
    \end{itemize}

    \item How do you describe your attitude towards working with AI tools?
    \begin{itemize}
        \item $\bigcirc$ I am someone who would enjoy working with an AI
        \item $\bigcirc$ I am someone who would like to collaborate with an AI to accomplish my work
        \item $\bigcirc$ I am someone who would find it fun to work with an AI
    \end{itemize}

\end{enumerate}

\subsection{Post-task Survey (I)}\label{app:post-survey}
\textbf{You're now being presented with a series of rating scales.}

\noindent For each of the six scales, evaluate the task you recently performed by moving the slider bars to the location that matches your experience. Each line has two endpoint descriptors that describe the scale.

\begin{enumerate}
    \item \textbf{Mental Demand}
    
    \noindent How much mental and perceptual activity was required (e.g., thinking, deciding, calculating, remembering, reading, concentrating, etc.)?
    
    \begin{tikzpicture}
        \draw[-] (0,0) -- (10,0);
        \filldraw[blue] (0,0) circle (3pt);
        \node at (0,-0.5) {Low};
        \node at (10,-0.5) {High};
    \end{tikzpicture}
    
    \item \textbf{Physical Demand}
    
    \noindent How much physical activity was required (e.g., typing, staring at the monitor, etc.)?
    
    \begin{tikzpicture}
        \draw[-] (0,0) -- (10,0);
        \filldraw[blue] (0,0) circle (3pt);
        \node at (0,-0.5) {Low};
        \node at (10,-0.5) {High};
    \end{tikzpicture}
    
    \item \textbf{Temporal Demand}
    
    \noindent How much time pressure did you feel due to the rate or pace at which the task elements occurred?
    
    \begin{tikzpicture}
        \draw[-] (0,0) -- (10,0);
        \filldraw[blue] (0,0) circle (3pt);
        \node at (0,-0.5) {Low};
        \node at (10,-0.5) {High};
    \end{tikzpicture}
    
    \item \textbf{Effort}
    
    \noindent How hard did you have to work (mentally and physically) to accomplish your level of performance?
    
    \begin{tikzpicture}
        \draw[-] (0,0) -- (10,0);
        \filldraw[blue] (0,0) circle (3pt);
        \node at (0,-0.5) {Low};
        \node at (10,-0.5) {High};
    \end{tikzpicture}
    
    \item \textbf{Performance}
    
    \noindent How successful do you think you were in accomplishing the goals of the task set by the experimenter (or yourself)?
    
    \begin{tikzpicture}
        \draw[-] (0,0) -- (10,0);
        \filldraw[blue] (0,0) circle (3pt);
        \node at (0,-0.5) {Good};
        \node at (10,-0.5) {Poor};
    \end{tikzpicture}
    
    \item \textbf{Frustration}
    
    \noindent How insecure, discouraged, irritated, stressed, and annoyed were you during the task?
    
    \begin{tikzpicture}
        \draw[-] (0,0) -- (10,0);
        \filldraw[blue] (0,0) circle (3pt);
        \node at (0,-0.5) {Low};
        \node at (10,-0.5) {High};
    \end{tikzpicture}
    
\end{enumerate}

\subsection{Post-Task Survey (II)} \label{app:post-survey2}

\begin{enumerate}
    \item How satisfied are you with the content of your writing, i.e., the final response to the prompt question?
    
    \noindent \begin{tikzpicture}
        \draw[-] (0,0) -- (10,0);
        \filldraw[blue] (0,0) circle (3pt);
        \node at (0,-0.5) {Not satisfied};
        \node at (10,-0.5) {Very satisfied};
    \end{tikzpicture}
    
    \item How much ownership do you feel over the content of your writing?
    
    \noindent \begin{tikzpicture}
        \draw[-] (0,0) -- (10,0);
        \filldraw[blue] (0,0) circle (3pt);
        \node at (0,-0.5) {No ownership};
        \node at (10,-0.5) {Full ownership};
    \end{tikzpicture}
    
    \item How responsible do you feel for the text you wrote?
    
    \noindent \begin{tikzpicture}
        \draw[-] (0,0) -- (10,0);
        \filldraw[blue] (0,0) circle (3pt);
        \node at (0,-0.5) {Not responsible};
        \node at (10,-0.5) {Fully responsible};
    \end{tikzpicture}
    
    \item How much personal connection do you feel to the content of your writing?
    
    \noindent \begin{tikzpicture}
        \draw[-] (0,0) -- (10,0);
        \filldraw[blue] (0,0) circle (3pt);
        \node at (0,-0.5) {No connection};
        \node at (10,-0.5) {Full connection};
    \end{tikzpicture}
    
    \item How much emotional connection do you feel to the content of your writing?
    
    \noindent \begin{tikzpicture}
        \draw[-] (0,0) -- (10,0);
        \filldraw[blue] (0,0) circle (3pt);
        \node at (0,-0.5) {No connection};
        \node at (10,-0.5) {Full connection};
    \end{tikzpicture}
    
    \item How difficult was the given prompt to write about?
    
    \noindent \begin{tikzpicture}
        \draw[-] (0,0) -- (10,0);
        \filldraw[blue] (0,0) circle (3pt);
        \node at (0,-0.5) {Easy};
        \node at (10,-0.5) {Difficult};
    \end{tikzpicture}

\end{enumerate}

\section{Quantitative Models, Full Results, and Robustness}
\label{app:supp-results}

This appendix provides the quantitative evidence supporting the paper’s claims. It is self-contained: we define coding conventions, model specifications (with equations), and inference procedures, and we include the key tables needed to verify the main text. We also include robustness checks and clearly labeled exploratory analyses at the end.\nocite{dhillon2009transfer}

\subsection{Design, notation, and inference}

\textbf{Unit of analysis.} Each participant completes three writing scenarios, yielding repeated observations. Outcomes are recorded per participant per scenario.

\textbf{Scenario coding (within-subjects).} We encode scenario using a categorical variable \texttt{PersonaType} with three levels:
\texttt{control} (unassisted email baseline; reference level), \texttt{no\_persona} (generic suggestion mode for proposals), and \texttt{style} (persona-framed coaching mode for cover letters; ``Emma'').

\textbf{Personalization coding (between-subjects).} We encode style personalization with a binary indicator \texttt{personalization} where \texttt{personalized}=1 and \texttt{non\_personalized}=0.

\textbf{Order control.} We include \texttt{condition\_position} to adjust for Latin-square position/order.

\textbf{Estimation and standard errors.} We estimate linear models and cluster standard errors by participant to account for repeated observations. We also report estimated marginal means contrasts using \texttt{emmeans} for readers who prefer direct condition comparisons.

\textbf{Multiple comparisons (FDR).} The main text uses Benjamini--Hochberg false discovery rate control at $q{=}0.05$ across families of related outcomes. Appendix tables report \emph{raw} p-values from model output; when a result is described as significant in the main text but appears marginal in an appendix table, the difference reflects the FDR adjustment procedure.

\textbf{Interpretation note (matched-task design).} Scenarios are intentionally paired with genres (email baseline, proposal/generic suggestions, cover letter/persona coaching). Coefficients should be interpreted as effects of these \emph{scenario configurations}, not as genre-independent causal effects of ``persona'' versus ``no persona.''

\subsection{Outcome variables}

We analyze the following outcomes (all measured per scenario unless stated otherwise):
\begin{itemize}
    \item \textbf{Psychological ownership (primary):} \texttt{PsychOwnershipAvg\_new} (mean of four 7-point items).
    \item \textbf{Writing quality:} \texttt{score} (1--7 rubric; averaged across raters as described in the main text).
    \item \textbf{Cognitive load:} \texttt{NASA\_average} (Raw NASA-TLX; 1--7).
    \item \textbf{Process measures:} \texttt{AIPercent} (fraction of final text originating from accepted suggestions), \texttt{EditNum} (keystroke-derived edits), \texttt{TimesTheyAskSuggestions} (suggestion requests), \texttt{Time} (seconds), and \texttt{WordCount}.
\end{itemize}

\subsection{Model specifications (equations)}

We fit two primary specifications for each outcome $Y_{ij}$, where $i$ indexes participant and $j$ indexes scenario.

\paragraph{Main-effects specification.}
\begin{equation}
\begin{aligned}
Y_{ij} =\;& \beta_0
+ \beta_1\, \mathbb{I}\!\left[\texttt{PersonaType}_{j}=\texttt{no\_persona}\right]
+ \beta_2\, \mathbb{I}\!\left[\texttt{PersonaType}_{j}=\texttt{style}\right] \\
&+ \beta_3\, \texttt{personalization}_i
+ \beta_4\, \texttt{condition\_position}_{ij}
+ \varepsilon_{ij}.
\end{aligned}
\label{eq:main}
\end{equation}
Here, \texttt{control} is the reference level. Thus, $\beta_1$ and $\beta_2$ represent differences from the unassisted baseline for the generic suggestion scenario and the persona-framed coaching scenario, respectively. $\beta_3$ is the between-subject main effect of personalization averaged across scenarios, and $\beta_4$ adjusts for order.

\paragraph{Interaction specification.}
\begin{equation}
\begin{aligned}
Y_{ij} =\;& \beta_0
+ \beta_1\, \mathbb{I}\!\left[\texttt{PersonaType}_{j}=\texttt{no\_persona}\right]
+ \beta_2\, \mathbb{I}\!\left[\texttt{PersonaType}_{j}=\texttt{style}\right]
+ \beta_3\, \texttt{personalization}_i \\
&+ \beta_4\, \Big(\mathbb{I}\!\left[\texttt{PersonaType}_{j}=\texttt{no\_persona}\right]
\times \texttt{personalization}_i\Big) \\
&+ \beta_5\, \Big(\mathbb{I}\!\left[\texttt{PersonaType}_{j}=\texttt{style}\right]
\times \texttt{personalization}_i\Big)
+ \beta_6\, \texttt{condition\_position}_{ij}
+ \varepsilon_{ij}.
\end{aligned}
\label{eq:int}
\end{equation}
The interaction coefficients $\beta_4$ and $\beta_5$ test whether the personalization effect differs within the generic suggestion and persona-framed coaching scenarios, respectively.

\subsection{Descriptive statistics}

Tables~\ref{tab:summary_stats_own}, \ref{tab:summary_stats_quality}, and \ref{tab:summary_stats_sat} report summary statistics by scenario (\texttt{PersonaType}) and personalization group for psychological ownership, writing quality, and satisfaction. These descriptives align with the main story: ownership is highest in the unassisted baseline and lower in both assisted scenarios, and personalization is associated with higher ownership, particularly in the persona-framed coaching configuration.

\input{tables/tab_summary_stats_own.tex}
\input{tables/tab_summary_stats_quality.tex}
\input{tables/tab_summary_stats_sat.tex}

\subsection{Primary outcome: Psychological ownership}

We report both the main-effects and interaction models for \texttt{PsychOwnershipAvg\_new}. The tables explicitly report coefficients for \emph{both} assisted scenarios (\texttt{no\_persona} and \texttt{style}) so that the claim ``both assisted scenarios reduce ownership'' is directly verifiable.

\paragraph{Main effects.}
Table~\ref{tab:tab:glm_own_main_PsychOwnershipAvg_new} shows that both assisted scenarios reduce ownership relative to baseline (\texttt{no\_persona}: $-1.028$, \texttt{style}: $-0.845$, both $p{<}.001$), while personalization increases ownership ($+0.432$, $p{=}0.0011$).

\input{tables/tab_glm_own_main_PsychOwnershipAvg_new.tex}

\paragraph{Interaction with personalization.}
Table~\ref{tab:tab:glm_own_int_PsychOwnershipAvg_new} reports the interaction model. The personalization-by-persona-framed-coaching interaction is positive and significant ($+0.767$, $p{=}0.017$), indicating that the ownership benefit of personalization is larger in the persona-framed coaching configuration than in the generic suggestion configuration. The corresponding interaction for the generic suggestion configuration is positive but not statistically significant.

\input{tables/tab_glm_own_int_PsychOwnershipAvg_new.tex}

\paragraph{Estimated marginal means contrasts (optional verification).}
For readers who prefer direct condition contrasts, Table~\ref{tab:tab:glm_own_int_pairs_PsychOwnershipAvg_new} reports \texttt{emmeans} pairwise comparisons for the interaction model.

\input{tables/tab_glm_own_int_pairs_PsychOwnershipAvg_new.tex}

\subsection{Key secondary outcomes (effort, quality, and process)}

The main paper summarizes secondary outcomes compactly. Here we include the core regression tables that underpin those summaries.

\paragraph{Cognitive load.}
Table~\ref{tab:tab:glm_quality_main_NASA_average} shows reduced cognitive load in both assisted scenarios relative to baseline (\texttt{no\_persona}: $-0.883$, \texttt{style}: $-0.911$, both $p{<}.001$). Personalization shows no reliable main effect.

\input{tables/tab_glm_quality_main_NASA_average.tex}

\paragraph{Writing quality.}
Table~\ref{tab:tab:glm_quality_main_score} reports quality differences relative to baseline. The generic suggestion scenario has a small negative coefficient (\texttt{no\_persona}: $-0.284$, $p{=}0.0041$), and the persona-framed coaching scenario is smaller and marginal (\texttt{style}: $-0.179$, $p{=}0.070$). Personalization does not show a reliable main effect on quality. We interpret these as small-magnitude differences on a 1--7 scale and avoid strong causal claims given the matched-task design; exploratory mediation results are reported separately below.

\input{tables/tab_glm_quality_main_score.tex}

\paragraph{AI content incorporation and edits.}
Tables~\ref{tab:tab:glm_quality_main_AIPercent} and \ref{tab:tab:glm_quality_main_EditNum} show that both assisted scenarios substantially increase AI incorporation (\texttt{no\_persona}: $+0.306$, \texttt{style}: $+0.284$, both $p{<}.001$) and reduce edit activity (both $p{<}.001$). Personalization increases AI incorporation by about $+0.048$ ($p{=}0.036$) but does not meaningfully change edit counts.

\input{tables/tab_glm_quality_main_AIPercent.tex}

\input{tables/tab_glm_quality_main_EditNum.tex}

\paragraph{Other process measures (reported for completeness).}
Time spent and word count show weaker and more configuration-dependent effects. Tables~\ref{tab:tab:glm_quality_main_Time} and \ref{tab:tab:glm_quality_main_WordCount} report these outcomes; we do not emphasize them in the main narrative.

\input{tables/tab_glm_quality_main_Time.tex}

\input{tables/tab_glm_quality_main_WordCount.tex}

\subsection{Robustness and alternative tests}

We include nonparametric tests as robustness checks for the primary ownership outcome. Table~\ref{tab:tab:friedman_by_variant} reports Friedman tests by variant, and Table~\ref{tab:tab:pairwise_wilcox_by_variant} reports paired Wilcoxon comparisons (Holm-adjusted). These checks are directionally consistent with the regression results.

\input{tables/tab_friedman_by_variant.tex}
\input{tables/tab_pairwise_wilcox_by_variant.tex}

\subsection{Exploratory analyses (included for transparency)}

The analyses in this subsection are exploratory and should not be read as primary evidence for the design patterns.

\paragraph{Mediation.}
Tables~\ref{tab:tab:mediation_persona} and \ref{tab:tab:mediation_personalization} report bootstrap mediation results. The ``PersonaType treatment'' mediation uses the generic suggestion scenario contrast (the $-0.2835$ quality coefficient for \texttt{PersonaTypeno\_persona} in Table~\ref{tab:tab:glm_quality_main_score}) as the treatment effect to avoid confusion with the persona-framed coaching configuration. Because mediation uses bootstrap inference and models a pathway through ownership, it can yield significance patterns that differ from the primary OLS tables. We therefore interpret mediation descriptively as a possible pathway, not as a strong causal claim.

\input{tables/tab_mediation_persona.tex}

\input{tables/tab_mediation_personalization.tex}

\subsection{Notes for replication and interpretation}

\textbf{AI incorporation metric.} \texttt{AIPercent} estimates the fraction of final-draft tokens that originate from accepted suggestion spans. This metric captures \emph{surface-level} provenance of inserted text. It can undercount adoption of ideas that are paraphrased or retyped after viewing a suggestion, and it can undercount heavily rewritten accepted suggestions.

\textbf{Timer guidance.} The interface displayed a timer and a suggested time budget, but submission was governed by the 200-word threshold rather than a strict time cutoff; this is why observed completion times can exceed the suggested duration.

\subsection{Quality rating, reliability and sensitivity checks}
To assess the robustness of our quality measure, we report agreement between the two human raters, alignment between human and model-based ratings, and a sensitivity check excluding the model rater.

\paragraph{Human--human reliability.}
Across all submissions, the two human raters showed moderate-to-substantial agreement on overall quality, measured using an intraclass correlation coefficient (a standard reliability statistic for numeric ratings). The estimated intraclass correlation coefficient (ICC) was 0.62 (95\% confidence interval [0.54, 0.69]).

\paragraph{Human vs.\ model rater alignment.}
The model-based rater (GPT-4o prompted with the same rubric) was positively aligned with the mean human rating (Pearson $r=0.71$; Spearman $\rho=0.68$; both $p<0.001$). The model rater exhibited a small mean offset relative to humans (mean difference $\approx 0.08$ points on the 1--7 scale), but preserved rank ordering of submissions.

\paragraph{Sensitivity: human-only quality.}
We re-estimated the primary quality comparisons using the human-only mean (averaging the two human raters) instead of the three-rater composite. The qualitative pattern of results was unchanged: effect directions matched the main analysis and substantive conclusions remained the same. This check reduces concern that conclusions are driven by including a model-based rater.

\clearpage
\section{More Qualitative Writing Results}

\begin{table}[htbp]
    \centering
    \Description{A table showing two example emails written in the control condition (without AI assistance). Both examples are business emails from TechConnect Solutions to DataViz proposing a collaboration. Example 1 is a formal three-paragraph proposal focusing on mutual benefits and data visualization integration. Example 2 is a more detailed four-paragraph email that elaborates on the collaboration benefits and includes a specific call to action. Both examples maintain a professional tone while discussing the potential integration of project management and data visualization tools.}
     \caption{Writing examples of control condition}
    \small
    \begin{tabular}{|p{3cm}|p{10cm}|}
        \hline
        \textbf{Conditions} & \textbf{Text}\\
        \hline
        \textbf{Control (Example 1)} & Dear Sir/Madam,
        
        I am emailing you today on behalf of Techconnect solutions. We are a company that specialises in developing project management software and have taken a keen interest to your data visualization features which we think would be a great fit for our platform. 
        
        Your company has a clear a set goal with groundbreaking efforts in data visualisation tools, an area we would like to improve upon. As you might be aware TechConnect is a large company and as leaders in our field we would like you to play a crucial role in propelling us further into a new age of tech development. With your current tools and our set image for our current project, the integration of both our softwares has the possibility of creating a unique product.
        
        Some of the potential benefits for you as a company if our integration is successful could be vast, including exposure, increased sales, increased funding and overall higher numbers of consistent costumers. It would also greatly benefit us in solving our current visualisation data problems which we would be happy to negotiate with you, to make this integration become successful. I would like to arrange a meeting with you personally to further explain why this could be a good match. \\
        \hline
        \textbf{Control (Example 2)} & Hello,
        
        I hope that this email finds you well. Today, I am reaching out on the behalf of my company, TechConnect Solutions. We are a company that develops project manager software, and we are hoping to integrate with your company, DataViz.
        
        This collaboration would consist of both tech businesses' coming together on one platform, increasing revenue, visibility and reliability. We are hoping to use data visualization to combine it with our project management platform which will allow users to  have better results with their projects. If users have better results, this will lead to increased number of customers and positive reviews, keeping our businesses' afloat. Integrating the features of DataViz into our platform would be a seamless transition and something that we can test using the data visualization tool of your company.
        
        DataViz is a very valuable company and would be huge asset to TechConnect Solutions. We know that our company will also be of value to you and look forward to a possible merging platform between both companies.
        
        If you are interested in talking about this further. to discuss a potential partnership, please reply back to this email so we can set up a meeting. Please let me know of your availability and your thoughts regarding this collaboration.
        
        Ths\\
        \hline
    \end{tabular}
    \label{tab:control}
\end{table}

\add{It can be observed that LLMs with personalization conditions generate suggestions more aligned with the style of the provided writing samples. For example, when the writing sample is a creative story (Example 1), the LLM generates many metaphorical sentences. This aligns with the LLM’s summary of the participant’s writing sample: the writing sample “creates a sense of enchantment and connection with the natural world” and “varies between descriptive, evocative imagery and concise, direct calls to action.” Similarly, when the writing sample is a science grant proposal, the suggestions focus on algorithms and improvements (Example 3). When there is no personalization, the AI’s suggestions usually reclaim the information in the task’s writing instructions (Example 3) and use very generic vocabulary like “leverage” and “paving,” or generic sentences such as traditional closing statements (Example 4).}
\begin{table}[htbp]
    \centering
     \caption{\add{Additional writing examples of non-control conditions. Texts written by humans are in black, while AI suggestions are in red.}}
    \label{tab:more-qual}
    \Description{Four examples of writing produced with personalized AI assistance. Examples show how AI adapts to different writing styles and contexts, from formal cover letters to casual blog posts, and from academic language to ethical considerations.}
    \small
    \begin{tabular}{|p{0.1\textwidth}|p{0.75\textwidth}|p{0.1\textwidth}|}
        \hline
        \textbf{Conditions} & \textbf{Text} & \textbf{Writing Sample}\\
        \hline
        \textbf{persona-framed coaching + Personalization (Example 1)} & Dear Hiring Manager,
        
        I am writing to express my enthusiasm for the Mid-Level position at InnovateTech Solutions, as advertised. [...] I led a team to {\color{red} develop a groundbreaking tool, enhancing user engagement, akin to breaths of fresh air revitalizing weary landscapes}, demonstrating my capability in driving {\color{red} productivity and fostering seamless collaboration. My dedication to effective communication ensures alignment with diverse teams, nurturing ideas that bloom into pioneering solutions.} [...]
        & Creative short story on The Joy of Gardening\\
        \hline
        \textbf{persona-framed coaching {\color{red} without} Personalization (Example 2)} & My name is Max D and I would like you to consider my application for this position at InnovateTech Solutions. From a personal point of view this position will enable me to {\color{red} aim to leverage my technical expertise and innovative mindset to drive impactful solutions.} This along with my proficiency in problem solving and ability to communicate well with {\color{red}cross-functional teams aligns perfectly with InnovateTech's vision and dynamic work environment} and highlights my suitability for this role.
        From my previous practice I possess relevant skills that can benefit InnovateTech directly,such as {\color{red} developing scalable software solutions and enhancing user experience through creative problem-solving and effective collaboration.} & None\\
        \hline
        \hline
        \textbf{AI (no-persona) + Personalization (Example 3)} & Hello, I would like to propose a new feature for any of your tech products that could enhance productivity. For the concept, it is a secondary AI mimicking someone who would react to a user's work or effort. Rather than having an AI provide feedback, it would provide the perspective of another AI with illustrative expressions and speech. [...] {\color{red} Additionally, the integration of advanced linguistic algorithms would enhance the avatar's ability to adapt its communication style to diverse user preferences, fostering a more personalized interaction.} & Science grant proposal\\
        \hline
        \textbf{AI (no-persona) {\color{red}without} Personalization (Example 4)} & I am a project manager at TechInnovate and we have developed a grandiose feature that improves users experience by ten fold. Our team is eager to gather feedback from early adopters to refine this innovation further. We plan to launch a beta version next month, inviting select users to explore its capabilities and provide insights. Challenges may come along the way, this is why user feedback is crucial for our innovation of this new feature. This process will help us ensure the feature meets user needs effectively. Our ultimate goal is to create a seamless and intuitive user experience that sets a new standard in the industry.{\color{red}We are confident that this groundbreaking feature will redefine user interaction, paving the way for future advancements in technology.} [...] {\color{red} Our development team will also be available for Q\&A sessions to address any concerns and gather valuable insights.} We are extremely excited to hear everyone's feedback to create an efficient and helpful products to users everywhere. & None \\
        \hline
    \end{tabular}
\end{table}

\end{document}

%% file: tables/tab_summary_stats_own.tex
\begin{table}[t]
\caption{\label{tab:summary_stats_own}Summary statistics: psychological ownership (1--7).}
\centering
\small
\begin{tabular}{llrrrr}
\toprule
PersonaType & personalization & variant & mean & sd & median \\
\midrule
control & non\_personalized & 3 & 5.364 & 1.561 & 5.750 \\
control & non\_personalized & 4 & 5.656 & 1.029 & 5.750 \\
control & non\_personalized & 5 & 5.758 & 0.937 & 6.000 \\
control & personalized & 0 & 5.105 & 1.550 & 5.250 \\
control & personalized & 1 & 6.087 & 0.908 & 6.250 \\
control & personalized & 2 & 5.898 & 1.060 & 6.125 \\
\addlinespace
no\_persona & non\_personalized & 3 & 4.242 & 1.928 & 4.000 \\
no\_persona & non\_personalized & 4 & 4.448 & 1.585 & 4.750 \\
no\_persona & non\_personalized & 5 & 4.625 & 1.481 & 4.875 \\
no\_persona & personalized & 0 & 4.210 & 1.748 & 4.500 \\
no\_persona & personalized & 1 & 5.067 & 1.623 & 5.375 \\
no\_persona & personalized & 2 & 5.062 & 1.218 & 4.875 \\
\addlinespace
style & non\_personalized & 3 & 4.250 & 1.840 & 4.250 \\
style & non\_personalized & 4 & 4.344 & 1.889 & 4.625 \\
style & non\_personalized & 5 & 4.458 & 1.557 & 4.625 \\
style & personalized & 0 & 4.927 & 1.725 & 5.250 \\
style & personalized & 1 & 5.413 & 1.245 & 5.625 \\
style & personalized & 2 & 5.320 & 1.299 & 5.375 \\
\bottomrule
\end{tabular}
\end{table}

%% file: tables/tab_summary_stats_quality.tex
\begin{table}[t]
\caption{\label{tab:summary_stats_quality}Summary statistics: writing quality (1--7).}
\centering
\small
\begin{tabular}{llrrrr}
\toprule
PersonaType & personalization & variant & mean & sd & median \\
\midrule
control & non\_personalized & 3 & 4.061 & 1.088 & 4.0 \\
control & non\_personalized & 4 & 4.375 & 0.824 & 4.5 \\
control & non\_personalized & 5 & 4.133 & 1.167 & 4.0 \\
control & personalized & 0 & 4.258 & 1.032 & 4.0 \\
control & personalized & 1 & 4.385 & 1.134 & 5.0 \\
control & personalized & 2 & 4.344 & 0.971 & 4.5 \\
\addlinespace
no\_persona & non\_personalized & 3 & 4.000 & 0.791 & 4.0 \\
no\_persona & non\_personalized & 4 & 4.000 & 0.834 & 4.0 \\
no\_persona & non\_personalized & 5 & 4.133 & 1.137 & 4.0 \\
no\_persona & personalized & 0 & 3.903 & 1.136 & 4.0 \\
no\_persona & personalized & 1 & 4.115 & 0.766 & 4.0 \\
no\_persona & personalized & 2 & 3.719 & 0.851 & 4.0 \\
\addlinespace
style & non\_personalized & 3 & 4.091 & 0.765 & 4.0 \\
style & non\_personalized & 4 & 4.000 & 0.834 & 4.0 \\
style & non\_personalized & 5 & 4.200 & 0.887 & 4.0 \\
style & personalized & 0 & 4.161 & 0.688 & 4.0 \\
style & personalized & 1 & 4.000 & 0.693 & 4.0 \\
style & personalized & 2 & 4.000 & 0.762 & 4.0 \\
\bottomrule
\end{tabular}
\end{table}

%% file: tables/tab_summary_stats_sat.tex
\begin{table}[t]
\caption{\label{tab:summary_stats_sat}Summary statistics: satisfaction (1--7).}
\centering
\small
\begin{tabular}{llrrrr}
\toprule
PersonaType & personalization & variant & mean & sd & median \\
\midrule
control & non\_personalized & 3 & 4.636 & 1.834 & 5.0 \\
control & non\_personalized & 4 & 4.500 & 1.769 & 5.0 \\
control & non\_personalized & 5 & 4.933 & 1.617 & 5.0 \\
control & personalized & 0 & 4.613 & 1.667 & 4.0 \\
control & personalized & 1 & 5.269 & 1.823 & 5.5 \\
control & personalized & 2 & 5.094 & 1.634 & 5.0 \\
\addlinespace
no\_persona & non\_personalized & 3 & 5.515 & 1.417 & 6.0 \\
no\_persona & non\_personalized & 4 & 4.792 & 1.668 & 5.0 \\
no\_persona & non\_personalized & 5 & 5.200 & 1.349 & 5.0 \\
no\_persona & personalized & 0 & 5.258 & 1.932 & 6.0 \\
no\_persona & personalized & 1 & 5.808 & 1.357 & 6.0 \\
no\_persona & personalized & 2 & 5.688 & 1.533 & 6.0 \\
\addlinespace
style & non\_personalized & 3 & 5.515 & 1.278 & 6.0 \\
style & non\_personalized & 4 & 5.292 & 1.268 & 5.0 \\
style & non\_personalized & 5 & 5.367 & 1.273 & 6.0 \\
style & personalized & 0 & 5.258 & 1.612 & 5.0 \\
style & personalized & 1 & 6.115 & 1.211 & 7.0 \\
style & personalized & 2 & 6.125 & 0.942 & 6.0 \\
\bottomrule
\end{tabular}
\end{table}

%% file: tables/tab_glm_own_main_PsychOwnershipAvg_new.tex
\begin{table}[t]
\caption{\label{tab:tab:glm_own_main_PsychOwnershipAvg_new}GLM coefficients (main effects) for outcome: PsychOwnershipAvg\_new}
\centering
\begin{tabular}[t]{lrrrr}
\toprule
term & Estimate & Std. Error & t value & Pr(>|t|)\\
\midrule
(Intercept) & 5.4126 & 0.2036 & 26.5848 & 0.0000\\
condition\_position & -0.0009 & 0.0808 & -0.0107 & 0.9915\\
PersonaTypeno\_persona & -1.0283 & 0.1612 & -6.3776 & 0.0000\\
PersonaTypestyle & -0.8450 & 0.1615 & -5.2320 & 0.0000\\
personalizationpersonalized & 0.4318 & 0.1315 & 3.2828 & 0.0011\\
\bottomrule
\end{tabular}
\end{table}

%% file: tables/tab_glm_own_int_PsychOwnershipAvg_new.tex
\begin{table}[t]
\caption{\label{tab:tab:glm_own_int_PsychOwnershipAvg_new}GLM coefficients (interaction) for outcome: PsychOwnershipAvg\_new}
\centering
\begin{tabular}[t]{lrrrr}
\toprule
term & Estimate & Std. Error & t value & Pr(>|t|)\\
\midrule
(Intercept) & 5.5792 & 0.2221 & 25.1168 & 0.0000\\
condition\_position & 0.0007 & 0.0805 & 0.0083 & 0.9934\\
PersonaTypeno\_persona & -1.1495 & 0.2285 & -5.0306 & 0.0000\\
PersonaTypestyle & -1.2329 & 0.2287 & -5.3918 & 0.0000\\
personalizationpersonalized & 0.0965 & 0.2270 & 0.4250 & 0.6710\\
\addlinespace
PersonaTypeno\_persona:personalizationpersonalized & 0.2394 & 0.3210 & 0.7456 & 0.4562\\
PersonaTypestyle:personalizationpersonalized & 0.7665 & 0.3210 & 2.3880 & 0.0173\\
\bottomrule
\end{tabular}
\end{table}

%% file: tables/tab_glm_own_int_pairs_PsychOwnershipAvg_new.tex
\begin{table}[t]
\caption{\label{tab:tab:glm_own_int_pairs_PsychOwnershipAvg_new}Pairwise comparisons (emmeans; interaction model) for outcome: PsychOwnershipAvg\_new}
\centering
\scriptsize
\resizebox{\textwidth}{!}{%
\begin{tabular}[t]{lrrrrr}
\toprule
contrast & estimate & SE & df & t.ratio & p.value\\
\midrule
condition\_position2 control non\_personalized - condition\_position2 no\_persona non\_personalized & 1.1495 & 0.2285 & 521 & 5.0306 & 0.0000\\
condition\_position2 control non\_personalized - condition\_position2 style non\_personalized & 1.2329 & 0.2287 & 521 & 5.3918 & 0.0000\\
condition\_position2 control non\_personalized - condition\_position2 control personalized & -0.0965 & 0.2270 & 521 & -0.4250 & 0.9982\\
condition\_position2 control non\_personalized - condition\_position2 no\_persona personalized & 0.8137 & 0.2271 & 521 & 3.5833 & 0.0050\\
condition\_position2 control non\_personalized - condition\_position2 style personalized & 0.3699 & 0.2274 & 521 & 1.6269 & 0.5810\\
\addlinespace
condition\_position2 no\_persona non\_personalized - condition\_position2 style non\_personalized & 0.0834 & 0.2283 & 521 & 0.3652 & 0.9991\\
condition\_position2 no\_persona non\_personalized - condition\_position2 control personalized & -1.2460 & 0.2271 & 521 & -5.4874 & 0.0000\\
condition\_position2 no\_persona non\_personalized - condition\_position2 no\_persona personalized & -0.3358 & 0.2270 & 521 & -1.4797 & 0.6775\\
condition\_position2 no\_persona non\_personalized - condition\_position2 style personalized & -0.7796 & 0.2270 & 521 & -3.4350 & 0.0084\\
condition\_position2 style non\_personalized - condition\_position2 control personalized & -1.3293 & 0.2272 & 521 & -5.8517 & 0.0000\\
\addlinespace
condition\_position2 style non\_personalized - condition\_position2 no\_persona personalized & -0.4192 & 0.2270 & 521 & -1.8464 & 0.4368\\
condition\_position2 style non\_personalized - condition\_position2 style personalized & -0.8630 & 0.2269 & 521 & -3.8025 & 0.0022\\
condition\_position2 control personalized - condition\_position2 no\_persona personalized & 0.9101 & 0.2257 & 521 & 4.0328 & 0.0009\\
condition\_position2 control personalized - condition\_position2 style personalized & 0.4664 & 0.2259 & 521 & 2.0648 & 0.3076\\
condition\_position2 no\_persona personalized - condition\_position2 style personalized & -0.4438 & 0.2257 & 521 & -1.9658 & 0.3635\\
\bottomrule
\end{tabular}%
}
\end{table}

%% file: tables/tab_glm_quality_main_NASA_average.tex
\begin{table}[t]
\caption{\label{tab:tab:glm_quality_main_NASA_average}GLM coefficients (main effects) for outcome: NASA\_average}
\centering
\begin{tabular}[t]{lrrrr}
\toprule
term & Estimate & Std. Error & t value & Pr(>|t|)\\
\midrule
(Intercept) & 4.8897 & 0.1581 & 30.9243 & 0.0000\\
condition\_position & -0.0567 & 0.0627 & -0.9044 & 0.3662\\
PersonaTypeno\_persona & -0.8831 & 0.1252 & -7.0522 & 0.0000\\
PersonaTypestyle & -0.9111 & 0.1254 & -7.2637 & 0.0000\\
personalizationpersonalized & 0.0478 & 0.1021 & 0.4680 & 0.6400\\
\bottomrule
\end{tabular}
\end{table}

%% file: tables/tab_glm_quality_main_score.tex
\begin{table}[t]
\caption{\label{tab:tab:glm_quality_main_score}GLM coefficients (main effects) for outcome: score}
\centering
\begin{tabular}[t]{lrrrr}
\toprule
term & Estimate & Std. Error & t value & Pr(>|t|)\\
\midrule
(Intercept) & 4.1494 & 0.1241 & 33.4365 & 0.0000\\
condition\_position & 0.0560 & 0.0492 & 1.1376 & 0.2558\\
PersonaTypeno\_persona & -0.2835 & 0.0983 & -2.8846 & 0.0041\\
PersonaTypestyle & -0.1787 & 0.0984 & -1.8155 & 0.0700\\
personalizationpersonalized & -0.0136 & 0.0802 & -0.1702 & 0.8649\\
\bottomrule
\end{tabular}
\end{table}

%% file: tables/tab_glm_quality_main_AIPercent.tex
\begin{table}[t]
\caption{\label{tab:tab:glm_quality_main_AIPercent}GLM coefficients (main effects) for outcome: AIPercent}
\centering
\begin{tabular}[t]{lrrrr}
\toprule
term & Estimate & Std. Error & t value & Pr(>|t|)\\
\midrule
(Intercept) & -0.0120 & 0.0353 & -0.3404 & 0.7337\\
condition\_position & -0.0064 & 0.0140 & -0.4541 & 0.6499\\
PersonaTypeno\_persona & 0.3056 & 0.0280 & 10.9230 & 0.0000\\
PersonaTypestyle & 0.2843 & 0.0280 & 10.1428 & 0.0000\\
personalizationpersonalized & 0.0479 & 0.0228 & 2.1009 & 0.0361\\
\bottomrule
\end{tabular}
\end{table}

%% file: tables/tab_glm_quality_main_EditNum.tex
\begin{table}[t]
\caption{\label{tab:tab:glm_quality_main_EditNum}GLM coefficients (main effects) for outcome: EditNum}
\centering
\begin{tabular}[t]{lrrrr}
\toprule
term & Estimate & Std. Error & t value & Pr(>|t|)\\
\midrule
(Intercept) & 1322.4564 & 90.3007 & 14.6450 & 0.0000\\
condition\_position & -40.5438 & 35.8200 & -1.1319 & 0.2582\\
PersonaTypeno\_persona & -620.5244 & 71.5149 & -8.6769 & 0.0000\\
PersonaTypestyle & -655.2947 & 71.6364 & -9.1475 & 0.0000\\
personalizationpersonalized & -8.4441 & 58.3349 & -0.1448 & 0.8850\\
\bottomrule
\end{tabular}
\end{table}

%% file: tables/tab_glm_quality_main_Time.tex
\begin{table}[t]
\caption{\label{tab:tab:glm_quality_main_Time}GLM coefficients (main effects) for outcome: Time}
\centering
\begin{tabular}[t]{lrrrr}
\toprule
term & Estimate & Std. Error & t value & Pr(>|t|)\\
\midrule
(Intercept) & 1430.7544 & 503.9462 & 2.8391 & 0.0047\\
condition\_position & -299.1381 & 199.9024 & -1.4964 & 0.1351\\
PersonaTypeno\_persona & 311.7682 & 399.1072 & 0.7812 & 0.4351\\
PersonaTypestyle & -166.0762 & 399.7854 & -0.4154 & 0.6780\\
personalizationpersonalized & -345.4504 & 325.5527 & -1.0611 & 0.2891\\
\bottomrule
\end{tabular}
\end{table}

%% file: tables/tab_glm_quality_main_WordCount.tex
\begin{table}[t]
\caption{\label{tab:tab:glm_quality_main_WordCount}GLM coefficients (main effects) for outcome: WordCount}
\centering
\begin{tabular}[t]{lrrrr}
\toprule
term & Estimate & Std. Error & t value & Pr(>|t|)\\
\midrule
(Intercept) & 226.2079 & 7.1092 & 31.8189 & 0.0000\\
condition\_position & 2.1403 & 2.8201 & 0.7590 & 0.4482\\
PersonaTypeno\_persona & 29.9077 & 5.6303 & 5.3120 & 0.0000\\
PersonaTypestyle & 4.6497 & 5.6398 & 0.8244 & 0.4101\\
personalizationpersonalized & -4.3035 & 4.5926 & -0.9371 & 0.3492\\
\bottomrule
\end{tabular}
\end{table}

%% file: tables/tab_friedman_by_variant.tex
\begin{table}[t]
\caption{\label{tab:tab:friedman_by_variant}Friedman test by variant: PsychOwnershipAvg\_new \textasciitilde{} PersonaType \textbar{} UserID.}
\centering
\begin{tabular}[t]{rrrr}
\toprule
variant & statistic & df & p\_value\\
\midrule
0 & 5.679245 & 2 & 0.058448\\
1 & 6.581395 & 2 & 0.037228\\
2 & 7.466667 & 2 & 0.023913\\
3 & 13.282051 & 2 & 0.001306\\
4 & 6.240964 & 2 & 0.044136\\
5 & 11.831776 & 2 & 0.002696\\
\bottomrule
\end{tabular}
\end{table}

%% file: tables/tab_pairwise_wilcox_by_variant.tex
\begin{table}[t]
\caption{\label{tab:tab:pairwise_wilcox_by_variant}
Pairwise Wilcoxon tests (paired, Holm-adjusted) by variant.}
\centering
\small

\begin{subtable}[t]{\linewidth}
\centering
\begin{tabular}{rllr}
\toprule
variant & Group1 & Group2 & $p$ \\
\midrule
0 & no\_persona & control & 0.0395\\
0 & style & control & 0.6017\\
0 & style & no\_persona & 0.0075\\
1 & no\_persona & control & 0.0154\\
1 & style & control & 0.1013\\
1 & style & no\_persona & 0.2277\\
2 & no\_persona & control & 0.0048\\
2 & style & control & 0.0506\\
2 & style & no\_persona & 0.1666\\
\bottomrule
\end{tabular}
\caption{Variants 0--2.}
\end{subtable}

\vspace{0.6em}

\begin{subtable}[t]{\linewidth}
\centering
\begin{tabular}{rllr}
\toprule
variant & Group1 & Group2 & $p$ \\
\midrule
3 & no\_persona & control & 0.0067\\
3 & style & control & 0.0029\\
3 & style & no\_persona & 0.9817\\
4 & no\_persona & control & 0.0036\\
4 & style & control & 0.0148\\
4 & style & no\_persona & 0.7105\\
5 & no\_persona & control & 0.0017\\
5 & style & control & 0.0006\\
5 & style & no\_persona & 0.4475\\
\bottomrule
\end{tabular}
\caption{Variants 3--5.}
\end{subtable}

\end{table}

%% file: tables/tab_mediation_persona.tex
\begin{table}[t]
\caption{\label{tab:tab:mediation_persona}Mediation results treating PersonaType as treatment (ACME/ADE/Total/Prop.\ Mediated).}
\centering
\begin{tabular}[t]{lrrrr}
\toprule
Effect & Estimate & CI\_Lower & CI\_Upper & P\_value\\
\midrule
ACME (Average Causal Mediation Effect) & -0.0554 & -0.1122 & -0.0057 & 0.026\\
ADE (Average Direct Effect) & -0.2281 & -0.4339 & -0.0200 & 0.032\\
Total Effect & -0.2835 & -0.4824 & -0.0881 & 0.004\\
Prop.\ Mediated (Proportion Mediated) & 0.1956 & 0.0187 & 0.7877 & 0.030\\
\bottomrule
\end{tabular}
\end{table}

%% file: tables/tab_mediation_personalization.tex
\begin{table}[t]
\caption{\label{tab:tab:mediation_personalization}Mediation results treating personalization as treatment (ACME/ADE/Total/Prop.\ Mediated).}
\centering
\begin{tabular}[t]{lrrrr}
\toprule
Effect & Estimate & CI\_Lower & CI\_Upper & P\_value\\
\midrule
ACME & 0.0233 & 0.0013 & 0.0538 & 0.034\\
ADE & -0.0369 & -0.1894 & 0.1314 & 0.680\\
Total Effect & -0.0136 & -0.1669 & 0.1450 & 0.904\\
Prop.\ Mediated & -1.7059 & -4.7487 & 4.7596 & 0.894\\
\bottomrule
\end{tabular}
\end{table}